\documentclass[aps,prd,twocolumn,showpacs,superscriptaddress,groupedaddress, nofootinbib]{revtex4}  
\usepackage{graphicx}    
\usepackage{dcolumn}    
\usepackage{bm}              
\usepackage{amssymb}   
\usepackage{subfigure}
\usepackage{amsmath}

\newcommand{\DZero}	      {D0}
\newcommand{\Lumi}              {$L$}
\newcommand{\IntLumi}          {$\cal{L}$}

\newcommand{\sigmaeff}       {$\sigma_{LM}$}
\newcommand{\sigmainel}     {$\sigma_{inel}$}

\newcommand{\pythia}             {\textsc{pythia}}
\newcommand{\pT}		        {$p_\mathrm{T}$}
\newcommand{\uncertainty}     {$\pm$2.0~mb}
\newcommand{\integrated}       {9.2}
\newcommand{\intUncertainty} {0.4~fb$^{-1}$}

\begin{document}

\hspace{5.2in} \mbox{FERMILAB -- TM -- 2529 -- E}

\title{The \DZero\ Run IIb Luminosity Measurement}

\affiliation{LPNHE, Universit\'es Paris VI and VII, CNRS/IN2P3, Paris, France}
\affiliation{Fermi National Accelerator Laboratory, Batavia, Illinois 60510, USA}
\affiliation{Northwestern University, Evanston, Illinois 60208, USA}
\affiliation{Michigan State University, East Lansing, Michigan 48824, USA}
\affiliation{University of Nebraska, Lincoln, Nebraska 68588, USA}
\affiliation{Brown University, Providence, Rhode Island 02912, USA}
\affiliation{Rice University, Houston, Texas 77005, USA}
\author{B.C.K.~Casey} \affiliation{Fermi National Accelerator Laboratory, Batavia, Illinois 60510, USA}
\author{M.~Corcoran} \affiliation{Rice University, Houston, Texas 77005, USA}
\author{K.~DeVaughan} \affiliation{University of Nebraska, Lincoln, Nebraska 68588, USA}
\author{Y.~Enari} \affiliation{LPNHE, Universit\'es Paris VI and VII, CNRS/IN2P3, Paris, France}
\author{E.~Gallas\footnote{now at Department of Physics, Oxford University, Oxford, United Kingdom}} \affiliation{Fermi National Accelerator Laboratory, Batavia, Illinois 60510, USA}
\author{I.~Katsanos} \affiliation{University of Nebraska, Lincoln, Nebraska 68588, USA}
\author{J.~Linnemann} \affiliation{Michigan State University, East Lansing, Michigan 48824, USA}
\author{J.~Orduna} \affiliation{Rice University, Houston, Texas 77005, USA}
\author{R.~Partridge\footnote{Visitor from SLAC National Accelerator Laboratory, Menlo Park, CA, USA}} \affiliation{Brown University, Providence, Rhode Island 02912, USA}
\author{M.~Prewitt} \affiliation{Rice University, Houston, Texas 77005, USA}
\author{H.~Schellman} \affiliation{Northwestern University, Evanston, Illinois 60208, USA}
\author{G.R.~Snow} \affiliation{University of Nebraska, Lincoln, Nebraska 68588, USA}
\author{M.~Verzocchi} \affiliation{Fermi National Accelerator Laboratory, Batavia, Illinois 60510, USA}

\date{ April 02, 2012 }

\begin{abstract}

An assessment of the recorded integrated luminosity is presented for data collected with the \DZero\ detector at the Fermilab Tevatron Collider from June 2006 to September 2011 (Run IIb). In addition, a measurement of the effective cross section for inelastic interactions, also referred to as the luminosity constant, is reported. This measurement incorporates new features that lead to a substantial improvement in the precision of the result. A luminosity constant of $\sigma_{LM} = 48.3 \pm 1.9 \pm 0.6$~mb is obtained, where the first uncertainty is due to the accuracy of the inelastic cross section used by both CDF and \DZero, and the second uncertainty is due to \DZero\ sources.  The recorded luminosity for the highest $E_T$ jet trigger is $\cal{L}_\mathrm{rec}$ = \integrated\ $\pm$ \intUncertainty, with a relative uncertainty of 4.3\%.

\end{abstract}

\pacs{13.85.Lg}
\maketitle


\section{\label{sec:intro}Introduction}
	An essential ingredient in cross section measurements is the integrated luminosity, \IntLumi, used to normalize the data sample. At \DZero, the instantaneous luminosity, \Lumi, is derived from hit rates produced from inelastic proton -- antiproton collisions registered in a dedicated detector system. Measured hit rates are converted to luminosity using a normalization procedure based on the total inelastic cross section, and the geometric acceptance and efficiency of the dedicated detector system for registering inelastic events. The measurement of the effective cross section for inelastic interactions, and the assessed recorded integrated luminosity for data collected with the \DZero\ detector at the Fermilab Tevatron Collider from June 2006 to September 2011 are reported. Luminosity \Lumi\ varied during that period in the range (5 -- 420)~$\mu \mathrm{b^{-1}s^{-1}}$ (equivalent to (5 -- 420)$\cdot 10^{30}~\mathrm{cm^{-2}s^{-1}}$). In this luminosity range, the average number of inelastic proton -- antiproton interactions per crossing ranges from 0.18 to 14.8 requiring an accurate treatment of  multiple interactions.

In this note, a short description of the detector used for the instantaneous luminosity measurement and of the data samples used for this study is followed by a review of the luminosity measurement technique. The following sections describe the backgrounds that affect the luminosity measurement, the calculation of the detector acceptance, and the calculation of the luminosity constant and its uncertainty. In the Appendices, the luminosity measurement technique and the background removal are described in more detail.

	\subsection{Luminosity Monitor Detector}
		\label{sub:lmdet}
\begin{figure}
	\centering
	\subfigure[~$r - z$ view of the two arrays]{\label{fig:lm-a} \includegraphics[scale=0.35]{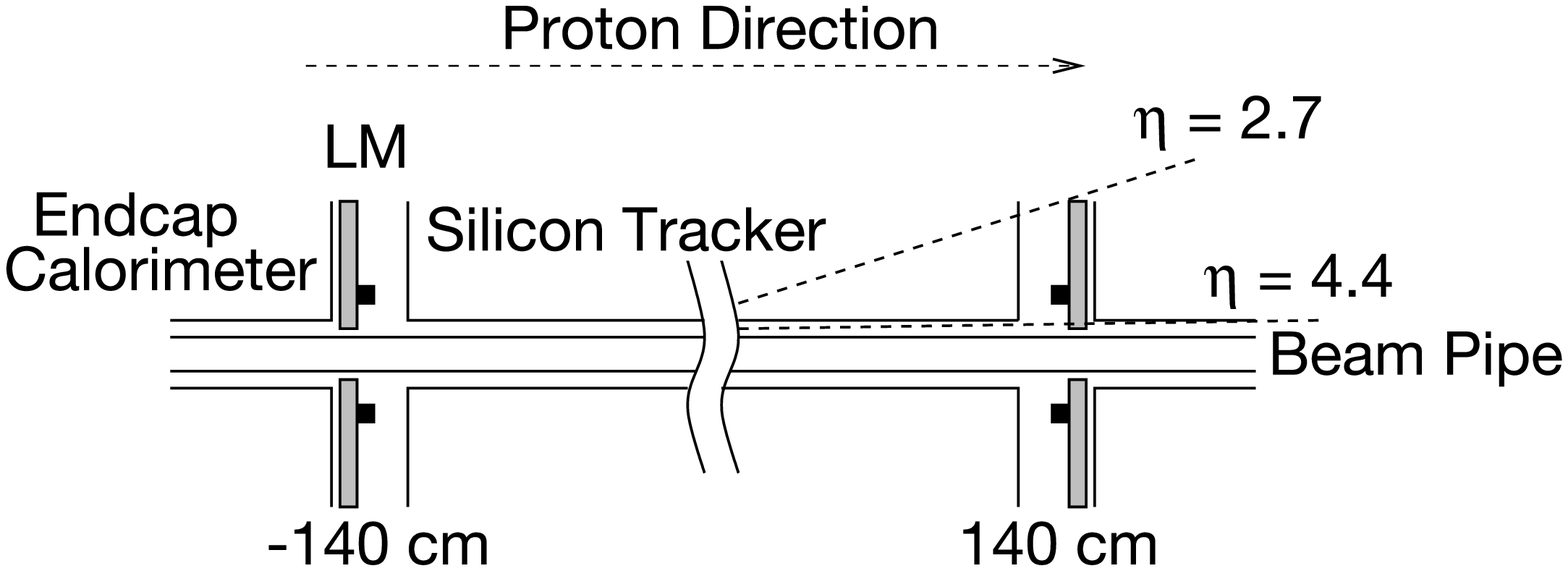}}
	\subfigure[~$r - \phi$ view of one array]{\label{fig:lm-b} \includegraphics[scale=0.35]{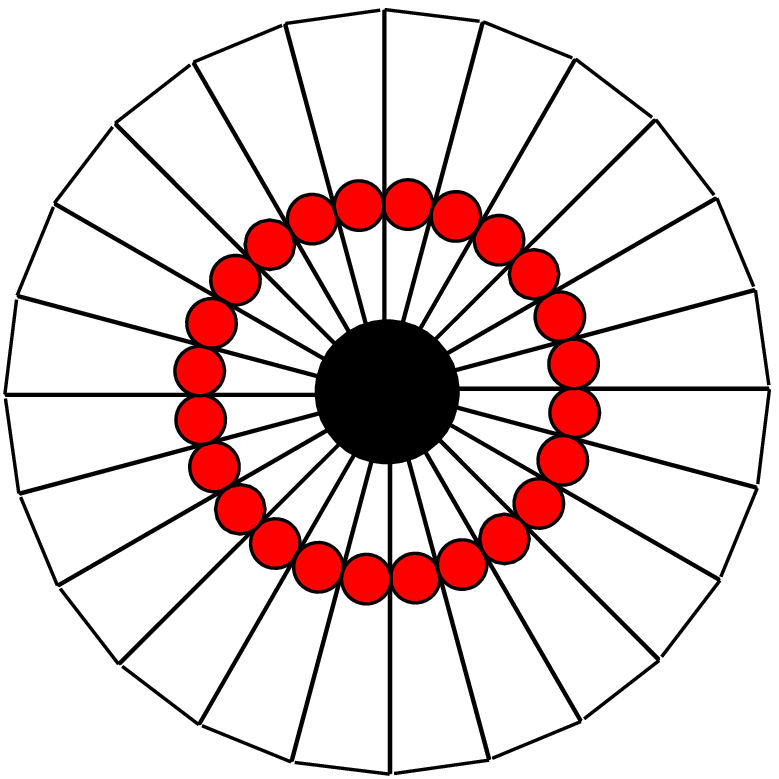}}
	\caption {The Luminosity Monitor layout. In (a) and (b) the solid dots represent the location of the PMTs.}
	\label{fig:lm}
\end{figure}

		The Luminosity Monitor (LM)~\cite{d0_lm, d0} consists of two arrays of scintillation counters mounted on the \DZero\ end-cap calorimeter cryostats as indicated in Fig.~\ref{fig:lm}. In the description of the \DZero\ detector a right-handed coordinate system is used. The $z$-axis is along the proton beam direction. The angles $\phi$ and $\theta$ are the azimuthal and polar angles, respectively. The $r$ coordinate denotes the perpendicular distance from the $z$ axis.

From the perspective of the proton beam, the upstream LM array is called the ``north'' LM and the downstream array is called the ``south'' LM.
Each array has 24 wedge-shaped scintillation counters with fine-mesh photomultiplier tube (PMT) readout. The PMT signals are amplified on the detector and are carried on low-loss cables to the LM VME electronics where the charge and the timing of PMT signals are measured.
Coverage is provided over the pseudorapidity interval of $2.7 < |\eta| < 4.4$, where $ \eta = \mathrm{-ln} \left [ \tan (\theta/2) \right ]$. 

The LM electronics identify in-time hits that are within $\pm6.4$~ns of the nominal time-of-flight from the center of the \DZero\ detector to the LM. This window is about three times the width of the time distribution for in-time hits. Halo particles typically produce hits that are $\sim$9~ns early in one of the detectors. A luminosity coincidence is identified when there is at least one in-time hit in both the north and the south LM detector arrays. Since beam crossings with many early hits from beam halo interactions can lead to luminosity measurement errors, a ``halo veto'' is applied when there are six or more early hits in one or both detector arrays. The beam crossings that do not trigger the halo veto  are called ``live crossings''\footnote[1]{The definition of live crossings in this context is with respect to the luminosity measurement and not the \DZero\ trigger system.}. 

	\subsection{Data Samples}
		\label{sub:datasamples}
		The data sample that \DZero\  recorded during Run II of the Fermilab Tevatron Collider is split in two periods: (i) data collected between April 2002 and February 2006 (Run IIa), and (ii) data collected from June 2006 to September 2011 (Run IIb). A major difference between the two periods is the addition of an inner silicon layer~\cite{Layer0}  to the \DZero\ Silicon Microstrip Tracker~\cite{d0smt} (SMT) during the 2006 shutdown. Other differences between the two periods include removal of a forward silicon disk on each end of the SMT and introduction of a new beryllium beam pipe with a flange near the LM.
The readout system of the LM detector was upgraded between Runs IIa and IIb to reduce the electronic noise~\cite{lm_elec}.

The data from the LM detector information includes measurements of the arrival time and pulse height information for each of the 48 LM counters. In addition, the LM electronics allow the accumulation of histograms of quantities calculated by the LM electronics for calibration and monitoring purposes. These histograms are accumulated at the beam crossing rate with no deadtime.

The Fermilab Tevatron Collider has 1113 possible radio frequency (RF) buckets. The minimal spacing between RF buckets where particles can be placed is one ``tick'' and corresponds to a gap of 132~ns. One turn of the Tevatron consists of 159 ticks, 
36 of which generally contain beam. The ticks that actually contain particles are called ``beam bunches'', and the collision of proton and anti-proton bunches is called a ``beam crossing'' or ``bunch crossing''. The beam bunches are arranged in 3 evenly spaced ``bunch trains'', separated by a 2.5~$\mu$s abort gap, and within each bunch train there are 12 beam bunches, each separated by 396~ns. Ticks that do not contain beam are referred to as ``empty ticks''. 

\begin{figure}
	\centering
	\includegraphics[scale=0.42]{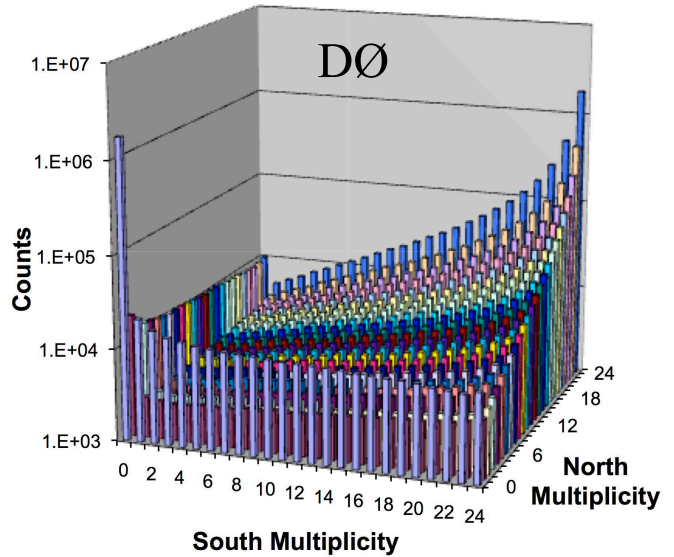}
	\caption {The 2D multiplicity distribution for live crossings, after background subtraction, collected with the histogramming feature of the LM electronics.}
	\label{fig:2d_multi}
\end{figure}

The LM electronics can accumulate two-dimensional (2D) distributions of the multiplicity of in-time hits for the north and south LM detectors. 
Figure~\ref{fig:2d_multi} shows an example distribution accumulated at a luminosity of 63~$\mu$b$^{-1}$s$^{-1}$ after background subtraction (see Section~\ref{sec:lmbkgd}). Three distinct components can be identified: (i) empty crossings with no LM hits, (ii) single-sided interactions where only one side has hits, and (iii) double-sided interactions where both sides have hits.

In this study extensive use is made of these 2D multiplicity distributions since they increase the number of events available for study by three orders of magnitude compared to an earlier analysis~\cite{runiia_lumi} (Run IIa) and provide the ability to measure the multiplicity distributions for a single bunch crossing, instead of averaging over 36 bunch crossings. Consequently, rigorous background subtraction techniques can be applied. In addition, data are now acquired over a short period of time ($\sim$8~mins total) such that the typical change in luminosity while the sample is acquired is less than 1\%. For contrast, the Run IIa analysis includes $\sim 1\%$ statistical errors due to the measurement being based in low statistics data samples (on the order of $\sim 10000$ beam crossings).

Histogram data samples were acquired over a period ranging from August 2008 to January 2009 for a variety of luminosities. In total, 35 such datasets were used for this study.

\section{\label{sec:overview}The \DZero\ luminosity measurement}
	The \DZero\ luminosity measurement is performed by counting the rate of north-south coincidences in the LM detectors using
\begin{equation}
	L = \frac{1}{\sigma_{LM}} \frac{dN}{dt},
\end {equation}
where \sigmaeff\ is the effective inelastic cross section for north-south coincidences as seen by the LM. We refer to the quantity $\sigma_{LM}$ as the ``luminosity constant''. The effective inelastic cross section is derived from the total inelastic cross section, \sigmainel , and adjusted for the LM system geometric acceptance and the efficiency for registering inelastic events. The inelastic cross section has been measured at the Tevatron by the E710, E811, and CDF experiments~\cite{cdf_xsection}. These experiments measure forward elastic scattering rates and use the optical theorem to determine the elastic, inelastic, and total $p \bar{p}$ cross section. A common averaging procedure~\cite{sigma_inel} for the E811 and CDF measurements has been adopted by the CDF and \DZero\ experiments, which yields an inelastic cross section of $\sigma_{inel} = 60.7 \pm 2.4$~mb at $\sqrt{s} = 1.96$~TeV.

The inelastic cross section can be subdivided into non-diffractive ($\sigma_{nd}$), single-diffractive ($\sigma_{sd}$), and double-diffractive ($\sigma_{dd}$) components. Single-diffractive collisions are characterized by having the proton (antiproton) diffractively disassociate into hadrons while the antiproton (proton) remains intact. As in elastic collisions, the momentum transfer is typically small, so that the intact antiproton (proton) exits the detector through the beam pipe. Double-diffractive collisions are similar to the single-diffractive collisions except that both the proton and the antiproton undergo diffractive disassociation. Like single-diffractive collisions, the particles produced tend to travel along the beam direction, thus producing large pseudorapidity gaps in the central region. Non-diffractive collisions represent the rest of the inelastic cross section and populate the full pseudorapidity region. Thus, the inelastic cross section can be expressed as

\begin{equation}
	\sigma_{inel} = \sigma_{nd} + \sigma_{sd} + \sigma_{dd}.
\end{equation}

The effective inelastic cross section, \sigmaeff, can be written as
\begin{equation}
	\begin{split}
		\sigma_{LM} =  \sigma_{inel} [  f_{nd}A_{nd} & + \left ( 1 - f_{nd} \right )f_{sd}A_{sd} \\
								         		& + \left ( 1-f_{nd} \right ) \left ( 1-f_{sd} \right ) A_{dd} ],
	\end{split}
	\label{eq:sigma_eff}
\end{equation}
where $f_{nd}$ is the fraction of the inelastic cross section attributed to the non-diffractive process and $f_{sd}$ is the fraction of the diffractive cross section attributed to the single-diffractive process, given by
\begin{align}
	{\label{eqn:non}} f_{nd}  & =  \frac{\sigma_{nd}}{\sigma_{inel}}, \\
	{\label{eqn:sd}}    f_{sd}  & =  \frac{\sigma_{sd}}{\sigma_{sd} + \sigma_{dd}}.
\end{align}		
The acceptances $A_{nd}$, $A_{sd}$, and $A_{dd}$ are the non-diffractive, single-diffractive, and double-diffractive acceptances, respectively, for producing at least one hit in both the north and south LM detectors.

A single-sided effective cross section $\sigma_N$ ($\sigma_S$) can be defined for producing hits in only the north (south) LM detector
\begin{equation}
	\begin{split}
		\sigma_{N}  = \sigma_{inel} [  f_{nd}A_{nd}^N &+ \left ( 1 - f_{nd} \right )f_{sd}A_{sd}^N \\
	& + \left ( 1-f_{nd} \right ) \left ( 1-f_{sd} \right ) A_{dd}^N ] ,
	\end{split}
	\label{eq:sigma_N}
\end{equation}

\begin{equation}
	\begin{split}
		\sigma_{S}  = \sigma_{inel} [  f_{nd}A_{nd}^S &+ \left ( 1 - f_{nd} \right )f_{sd}A_{sd}^S \\
	& + \left ( 1-f_{nd} \right ) \left ( 1-f_{sd} \right ) A_{dd}^S ] ,
	\end{split}
	\label{eq:sigma_S}
\end{equation}
where $A_{nd}^{N(S)}$, $A_{sd}^{N(S)}$, $A_{dd}^{N(S)}$ are the non-diffractive, single-diffractive, and double-diffractive acceptances, respectively, for producing at least one hit in the north (south) LM detector and no hits in the south (north) LM detector.
Earlier analyses of the luminosity constant~\cite{runiia_lumi} treated the north and south single-sided effective cross sections as being equal. In this analysis, small differences are found in both data and Monte Carlo (MC) simulations. These differences are attributed to asymmetries of the \DZero\ detector (e.g., the north endcap calorimeter is $\sim 4$~cm closer to the $p \bar{p}$ interaction point than the south endcap calorimeter). Consequently, a separate calculation of the north and south single-sided acceptances and cross sections is performed.

The ``empty crossing method'', which accounts for multiple interactions in a beam crossing, is used to measure the \DZero\ luminosity (see Appendix~\ref{app:empty}). Poisson statistics is used to relate the luminosity to the probability that in a beam crossing there is not a north-south coincidence. For a beam crossing not to have a north-south coincidence there must be no double-sided interactions characterized by $\sigma_{LM}$. In addition, there should be no pile-up within a single beam crossing of two single-sided interactions producing hits in both the north and south sides of the LM.

The probability that a beam crossing has no north-south coincidences, $P(0)$, is given by
\begin{equation}
	\label{eq:poisson}
	\begin{split}
	& P \left ( 0 \right ) = \\
	& e^{-\sigma_{LM} L / f} \left ( e^{-\sigma_N L / f} + e^{-\sigma_S L / f} - e^{-(\sigma_N + \sigma_S) L / f} \right ) ,
	\end{split}
\end{equation}
where $L$ is the luminosity and $f$ is the beam crossing frequency. The first factor is the probability for having no $p \bar{p}$ interactions giving a north-south coincidence. The term in the parenthesis corrects for two single-sided interactions in the same beam crossing mimicking a north--south coincidence.

The rate of live crossings with in-time hits in both the north and south LM detectors, $R_{LM}$, and the live crossing rate, $R_{Live}$, are measured in data. The probability for an empty beam crossing is derived from these rates to be
\begin{equation}
	P \left ( 0 \right ) = 1 - \frac{R_{LM}}{R_{Live}} .
\end{equation}

Given $P(0)$, Eq.~\ref{eq:poisson} is solved for the luminosity $L$, making use of the effective cross section \sigmaeff\ and the single-sided cross sections $\sigma_N$ and $\sigma_S$. This calculation is performed for each of the 36 beam bunches independently since the luminosity, and thus $P(0)$, is different for each bunch. More details about the empty crossing method and its application to high luminosities, where the average number of $p \bar{p}$ interactions per beam crossing can exceed 14, are given in Appendix~\ref{app:empty}.

\section{\label{sec:lmbkgd}Luminosity Monitor Backgrounds}
	The LM is sensitive to two types of backgrounds: (i) out-of-time, and (ii) beam halo backgrounds. The out-of-time background is characterized by hits with an approximately uniform arrival time distribution for the $\sim$40~ns measurement window before the beam crossing, with a significant variation in background rate over the 12 beam bunches in a bunch train. These hits randomly occur within the LM timing window, giving rise to in-time background hits. Studies of the out-of-time background over the entire revolution cycle of the Tevatron are presented in Appendix~\ref{app:background}. These studies show that the rate of background hits is proportional to the \DZero\ luminosity, indicating that they are due to secondary particles from beam-beam interactions in previous beam crossings. This background is found to have an effective single-sided cross section of $0.9 \pm 0.1$~mb for both the north and south sides. It is assumed to originate mainly from low energy neutrons that interact in the LM scintillator, but it is possible that other sources, such as short-lived activation products, contribute.

Beam halo backgrounds occur when a proton or antiproton leaves the beam pipe upstream of the interaction point and produces secondary particles that are detected by the LM. Beam halo typically produces a shower in the upstream calorimeter that hits the upstream LM, and continues to the downstream LM. Typically, the upstream LM counters will have out-of-time hits that arrive $\sim$9~ns earlier than particles from beam-beam collisions, while the downstream counters will have in-time hits that arrive at approximately the same time as particles from beam-beam collisions. Beam crossings with six or more early hits are vetoed in the luminosity calculation causing $\sim$1\% of the beam crossings to be removed. The luminosity calculation is not sensitive to the selection of the veto value.

During normal operation, the beam halo backgrounds are a few percent of the out-of-time background (e.g. for a luminosity of $100~\mu \mathrm{ b^{-1} s^{-1}}$ with a background cross section of 0.9~mb there is a 90~kHz background rate, whereas the typical halo rates are in the range of a few kHz) and have a negligible effect on the luminosity measurement. The remainder of this section focuses on describing and removing the out-of-time background.

Table~\ref{tab:bkgd_multi} shows a typical high luminosity background rates distribution measured in an empty tick immediately before a beam crossing. One or more background hits is present in 21\% of these empty ticks, and 1.3\% of them have hits in both north and south LM detectors. 

\begin{table}
\caption{Background multiplicity rates distribution measured in an empty tick immediately before a beam crossing. The data were acquired at a luminosity of $272~\mu \mathrm{ b^{-1} s^{-1}}$. The rows indicate the north multiplicity rate and the columns the south multiplicity rate, where the rates shown are normalized to the total number of events (5756304 events).}
	\begin{ruledtabular}
		\begin{tabular}{c|ccccc}
			N/S      & 0       & 1       & 2        & 3        & $\geq 4 $    \\  \hline
			0        & 79\%    & 8.1\%   & 1.3\%    & 0.18\%   & 0.06\%       \\
			1        & 9\%     & 0.9\%   & 0.15\%   & 0.022\%  & 0.007\%      \\
			2        & 1.5\%   & 0.18\%  & 0.027\%  & 0.004\%  & 0.0011\%      \\
			3        & 0.20\%  & 0.023\% & 0.005\%  & 0.0006\% & 0.0003\%      \\
			$\geq 4$ & 0.06\%  & 0.007\% & 0.0014\% & 0.0004\% & 0.0002\%      \\
		\end{tabular}
	\end{ruledtabular}
\label{tab:bkgd_multi}
\end{table}

The largest contribution of the background in the determination of \sigmaeff\ occurs when an empty crossing is converted into a single-sided crossing with one background hit. Previous determinations of the \sigmaeff~\cite{runiia_lumi} have used two techniques for removing the background contribution to the hit multiplicity distribution. The first technique used sidebands in the arrival time distribution to estimate the out-of-time background in events with a single hit in the detector and make the appropriate subtraction. The second technique required at least two hits (instead of one) on the opposite side when determining the non-diffractive fraction. 

For this study a different technique is used, for which a detailed description can be found in Appendix~\ref{app:bkgd_unfolding}. If signal and background hits are uncorrelated, the probability $D_{ij}$ of having $i$ north counters and $j$ south counters with observed hits is given by
\begin{equation}
	D_{ij}  = \sum_{\substack{l = 0 \\ m = 0}}^{\substack{l \leq i \\ m \leq j}}
		      \sum_{\substack{p = 0 \\ q = 0}}^{\substack{p \leq i \\ q \leq j}}
		      S_{lm} B_{pq} f_{lpi} f_{mqj} \Theta \left (l + p -i \right ) \Theta \left (m + q -j \right ),
\end{equation}
where $S_{lm}$ is the probability for having $l$ north counters and $m$ south counters with signal hits, $B_{pq}$ is the probability for having $p$ north counters and $q$ south counters with background hits, and $\Theta$ is the Heaviside step function
\begin{equation}
		\Theta \left ( l + p - i \right ) = 
		\begin{cases}
			 1 & \mathrm{~for~} l + p \geq i \\
			 0 & \mathrm{~for~} l + p < i.
		\end{cases} 
\end{equation}
 The combinatoric factor $f_{lpi}$ (and similarly $f_{mqj}$) represents the probability for observing $i$ north counters with hits given that there were $l$ north counters with signal hits and $p$ north counters with background hits. The factor $f_{lpi}$ is given by
\begin{equation}
	f_{lpi} = \frac{l! \left ( N - l \right ) ! p! \left ( N - p \right )!}{ \left ( l + p - i \right )! \left ( i - p \right ) ! \left ( i - l \right ) ! \left ( N - i \right ) ! N!}, 
\end{equation}
where $N = 24$ is the number of counters on a side.

The observed signal and background probability distribution can be obtained from the multiplicity histogram associated with a beam crossing, and the background probability distribution can be obtained from the multiplicity histogram obtained from the empty tick immediately before the beam crossing.   Thus, we can solve the above set of linear equations for the background-free signal probability distribution.

\begin{figure}
	\centering
	\includegraphics[scale=0.47]{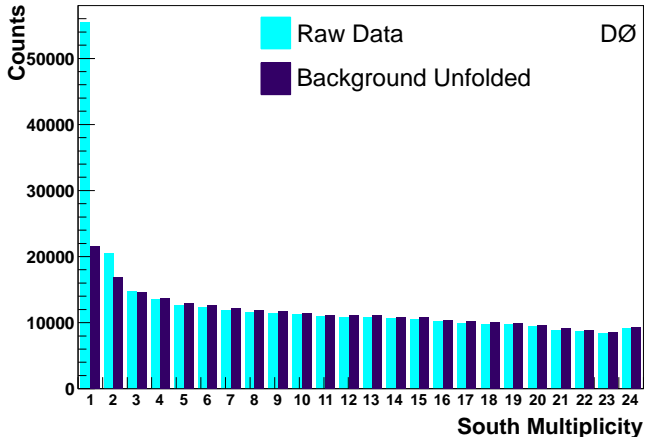}
	\caption {Example of the effect of background subtraction. The south multiplicity distribution is plotted for beam crossings where there are no hits in the north LM detector. The ``Raw Data'' histogram shows the multiplicity distribution before background subtraction.}
	\label{fig:south_multi}
\end{figure}

The effect of this background unfolding procedure is best illustrated by looking at its effect on beam crossings where one side has no hits. Figure~\ref{fig:south_multi} shows a slice of a 2D multiplicity distribution taken at a luminosity of $63~\mu \mathrm{b^{-1} s^{-1}}$. The slice plotted  shows the south multiplicity when there are no hits in the north LM detector. The prominent peak for a single south hit is substantially reduced after background subtraction, as would be expected from the shape of the background multiplicity distribution in Table~\ref{tab:bkgd_multi}. 

The background unfolding is performed on the 2D multiplicity distribution, and it has certain features that do not appear in a 1D unfolding procedure. For example, in Fig.~\ref{fig:south_multi} the number of entries in the higher multiplicity bins is larger after the background unfolding procedure than for the raw data. In the 1D unfolding procedure, entries can only migrate from higher multiplicity to lower multiplicity. In a 2D unfolding procedure, entries can also migrate between 1D slices. In the example shown in Fig.~\ref{fig:south_multi}, diffractive events with no north signal hits but $> 0$ north background hits are not included in the raw data (since the slice with no observed north hits is examined), but are included in the background subtracted measurement since these events truly have no north signal hits. The net effect of these two migrations is that bins with more than 3 south hits have more entries after background subtraction than before.

\section{\label{sec:lmconstant}Luminosity Constant Determination}
	\subsection{\label{subsec:acceptance}Luminosity Monitor Acceptance Calculation}
		The LM acceptances are calculated by simulating non-diffractive, single-diffractive, and double-diffractive events in the \DZero\ detector. Simulated events for each of the three subprocesses are generated using \pythia~\cite{pythia}, with the CTEQ6L1~\cite{cteq} parametrization of the parton distribution functions (PDFs), and utilizing the ``Tune A'' parameter set~\cite{rick_field} that is optimized to reproduce CDF data. The longitudinal distribution of the $p \bar{p}$ interaction vertex is generated as a Gaussian with an RMS of $\sim25$~cm.  In the transverse directions the beam is generated as a Gaussian centered at the origin with a width of about 100~$\mu$m. These Monte Carlo events are then processed through the standard \DZero\ detector simulation based on {\textsc{geant3}}~\cite{geant}, using a modified detector geometry with the material model adjusted to match the multiplicity distribution in the LM as observed in the data.

\begin{figure}
	\centering
	\includegraphics[scale=0.47]{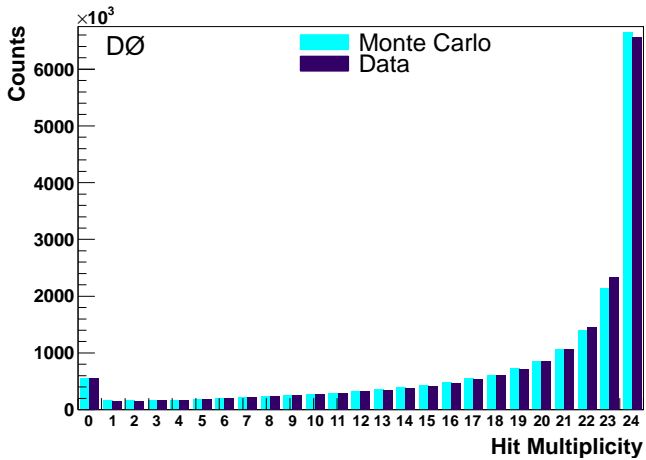}
	\caption {Hit multiplicity when there are one or more hits on the opposite side. The north and south distributions have been
	combined to reduce MC statistical errors. The data are from a background subtracted multiplicity histogram taken at a luminosity
	of 63~$\mu \mathrm{b^{-1} s^{-1}}$. The MC samples represent the tuned material model.}
	\label{fig:dataMCcomp1}
\end{figure}

\begin{figure}
	\centering
	\includegraphics[scale=0.47]{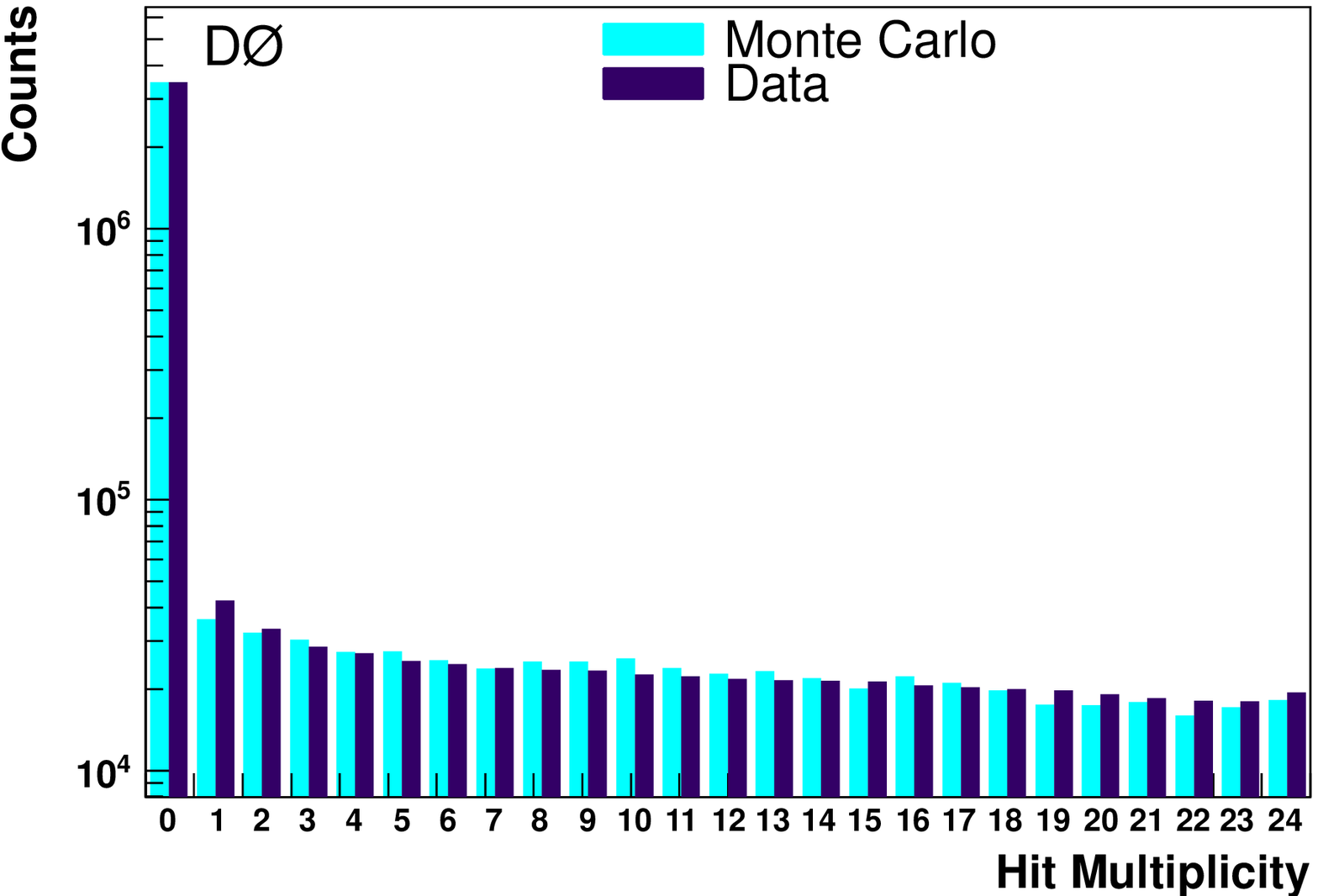}
	\caption {Hit multiplicity when there are no hits on the opposite side. The north and south distributions have been
	combined to reduce MC statistical errors. The data are from a background subtracted multiplicity histogram taken at a luminosity
	of 63~$\mu \mathrm{b^{-1} s^{-1}}$. The MC samples represent the tuned material model.}
	\label{fig:dataMCcomp2}
\end{figure}

As a check that the tuned material model reproduces the hit multiplicity histograms, Figs.~\ref{fig:dataMCcomp1} and~\ref{fig:dataMCcomp2} show the hit multiplicity for a histogram acquired at a luminosity of 63~$\mu \mathrm{b^{-1} s^{-1}}$. 
The data points have the out-of-time background contribution removed using the background unfolding procedure described in Appendix~\ref{app:bkgd_unfolding}. In Fig.~\ref{fig:dataMCcomp1} at least one opposite side hit is required, while in Fig.~\ref{fig:dataMCcomp2} the requirement is that there are no opposite side hits. The MC distributions were generated using the non-diffractive and single-diffractive fractions used in the final luminosity constant as discussed below. 
The hit multiplicity distributions for these data and MC samples are observed to be in good agreement.

The number of hits that satisfy the timing criteria and have charge above threshold is counted separately for the north and south LM detectors, with the MC charge threshold and timing resolution adjusted to reproduce the data. Events are then classified into one of the following geometrical categories: (i) events that have at least one hit in both the north and south LM detectors (NS), (ii) events that have at least one hit in the north detector and no hits in the south detector (N Only), (iii) events that have at least one hit in the south detector and no hits in the north detector (S Only) and (iv) events that have no hits in either detector (Empty). The fraction of events in each category determines the associated acceptance shown in Table~\ref{tab:acceptances}.
\begin{table*}
\caption{LM acceptances and their statistical uncertainties, for each type of inelastic process and geometrical category.}
	\begin{ruledtabular}
		\begin{tabular}{cccc}
			Category & Non-Diffractive & Single-Diffractive & Double-Diffractive \\
					&        Acceptance   &         Acceptance       &      Acceptance \\ \hline
			NS          & $0.9924 \pm 0.0009$ & $0.326 \pm 0.005$ & $0.500 \pm 0.005$ \\
			N Only   & $0.0048 \pm 0.0007$ & $0.224 \pm 0.004$ & $0.203 \pm 0.004$ \\
			S Only   & $ 0.0026 \pm 0.0005$ & $0.225 \pm 0.004$ & $0.212 \pm 0.004$ \\
			Empty    & $ 0.0002 \pm 0.0001$ & $ 0.225 \pm 0.004$ & $ 0.0857 \pm 0.0028$ \\	
		\end{tabular}
	\end{ruledtabular}
\label{tab:acceptances}
\end{table*}
The ``N Only'' and ``S Only'' acceptances are not identical since the \DZero\ detector, and its MC description, is not north/south symmetric.

	\subsection{\label{subsec:nondiff_fraction}Determination of the Non-Diffractive Fraction}
		The non-diffractive fraction can be related to the ``zero fraction'' measured in the data by using Poisson statistics. ``Zero fraction'' is the fraction of beam crossings that have no hits on a given side (north or south) when there is at least one hit on the opposite side.

Starting with the north zero fraction, the probability of having no two-sided interactions and no one-sided interactions hitting the north side is
\begin{equation}
 	P \left ( N = 0 \right ) = e^{- \left ( \sigma_{LM} + \sigma_N \right ) L / f}.
\end{equation} 
Having zero hits on the north side while having at least one hit on the south side has a probability
\begin{equation}
 	P \left ( N = 0, S > 0 \right ) = e^{- \left ( \sigma_{LM} + \sigma_N \right ) L / f} \left ( 1 - e^{- \sigma_S L / f} \right ).
\end{equation} 
Thus the north zero fraction is given by:
\begin{equation}
	\label{eq:zero_fraction}
	\begin{split}
		f_0^N & = \frac{P \left ( N = 0, S > 0 \right )}{P \left ( S > 0 \right )} =  \frac{P \left ( N = 0, S > 0 \right )}{1 - P \left ( S = 0 \right )} \\
			   & = \frac{e^{- \left ( \sigma_{LM} + \sigma_N \right ) L / f} \left ( 1 - e^{- \sigma_S L / f} \right )} 
			   	         {1 - e^{- \left ( \sigma_{LM} + \sigma_S \right ) L / f}}.
	\end{split}
\end{equation}
The south zero fraction is obtained similarly, by exchanging $\sigma_N$ with $\sigma_S$.

The cross sections in the above equations depend on the total inelastic cross section \sigmainel, the non-diffractive fraction $f_{nd}$, the single-diffractive fraction $f_{sd}$, and the LM acceptances (see Eqs.~\ref{eq:sigma_eff},~\ref{eq:sigma_N},~\ref{eq:sigma_S}). The single-diffractive fraction is taken to be $f_{sd} = 0.57 \pm 0.21$~\cite{f_sd}.

Having already evaluated the LM acceptances, the only quantity still needed is the luminosity for the data sample. In a given sample, the luminosity can be determined from the measured LM coincidence probability
\begin{equation}
	\label{eq:lumi_nd}
	P \left ( N > 0, S > 0 \right ) = 1 - P \left ( 0 \right ),
\end{equation}
where $P(0)$ is given by Eq.~\ref{eq:poisson}. Equations~\ref{eq:zero_fraction} and~\ref{eq:lumi_nd} are used to solve for the two remaining unknowns: the non-diffractive fraction $f_{nd}$, and the luminosity $L$. 

\begin{figure}
	\centering
	\includegraphics[scale=0.47]{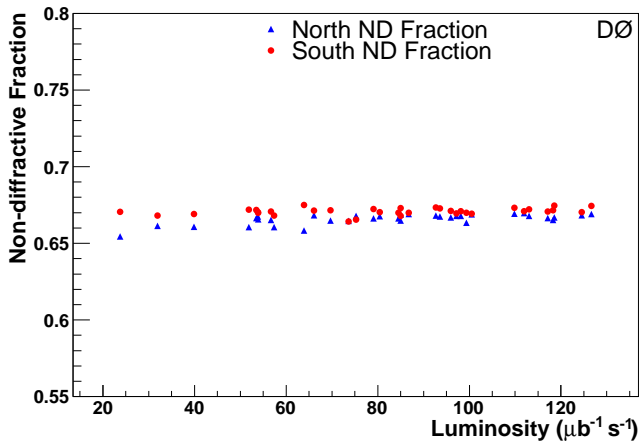}
	\caption {Non-diffractive fraction measurements as a function of luminosity. Separate fits are performed to the north and south zero fractions.}
	\label{fig:fnd}
\end{figure}

In the histogram data sample considered, 35 multiplicity histograms are acquired for the first beam bunch in the first beam train. For each of these histograms, the north and south zero fractions are calculated and the non-diffractive fraction determined. Results can be seen in Fig.~\ref{fig:fnd}, where the luminosity of the first beam bunch of the first beam train is extrapolated to the total delivered luminosity assuming that each beam bunch has the same luminosity. The statistical error on the non-diffractive fraction measurement ranges from 0.0006 to 0.0019 and is negligible.

The north and south non-diffractive fractions are similar but exhibit some systematic differences. The north non-diffractive fraction has a higher dispersion, and exhibits a small correlation with luminosity which is attributed to the effects of beam halo, where the rate for proton halo is in most cases substantially larger than for antiproton halo. Proton halo creates out-of-time hits in the north LM detector and in-time hits in the south LM detector. If a proton beam halo event occurs during an otherwise empty crossing, it will be counted as a single-sided event, thus increasing the north zero fraction. Since the diffractive processes are much more likely to have single-sided events than the non-diffractive processes,  an increased zero fraction corresponds to a decrease in the non-diffractive fraction. The fraction of empty beam crossings decreases  rapidly with increasing luminosity, so that the effect of beam halo is largest at low luminosity, which is consistent with the observed trend. The observed north/south difference is considered as a systematic uncertainty in the calculation of \sigmaeff.
Averaging the north and the south non-diffractive fractions for the 35 luminosity points considered results in $ \langle f_{nd} \rangle = 0.668$ with an RMS spread of 0.002.

	\subsection{\label{subsec:lmconstant}Luminosity Constant}
\begin{figure}
	\centering
	\includegraphics[scale=0.47]{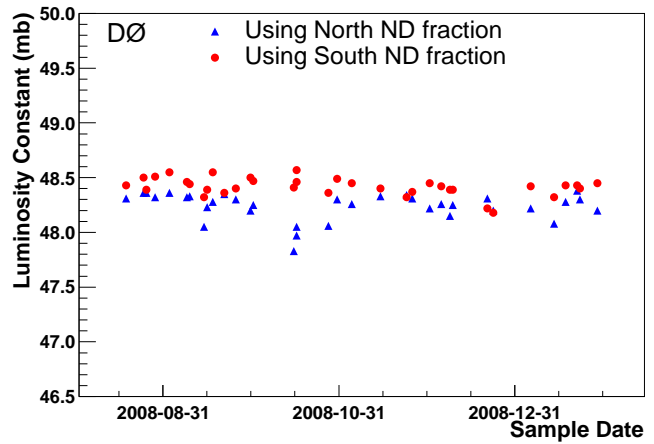}
	\caption {Measured luminosity constant as a function of time.}
	\label{fig:cst_time}
\end{figure}

\begin{figure}
	\centering		
	\includegraphics[scale=0.47]{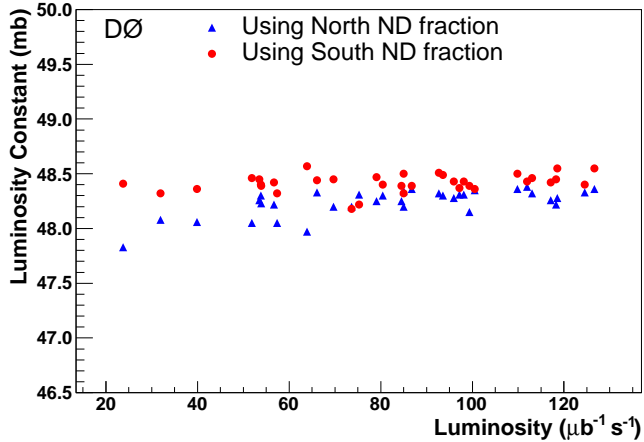}
	\caption {Measured luminosity constant as a function of the \DZero\ luminosity.}
	\label{fig:cst_lumi}
\end{figure}

		The calculated LM acceptances (Table~\ref{tab:acceptances}) and the non-diffractive fraction (evaluated in section~\ref{subsec:nondiff_fraction}), 
together with $\sigma_{inel} = 60.7 \pm 2.4$~mb and $f_{sd} = 0.57 \pm 0.21$ are used to determine the luminosity constant, \sigmaeff, based on Eq.~\ref{eq:sigma_eff}, for each data sample. Figure~\ref{fig:cst_time} shows the consistency of the measurement of \sigmaeff\ during the period of $\approx$6 months when the multiplicity histograms were collected. 
The luminosity constant is labeled as ``North'' when the north non-diffractive fraction has been used, while it is labeled as ``South'' when the south non-diffractive fraction has been used.
Figure~\ref{fig:cst_lumi} shows the luminosity dependence of the \sigmaeff\ determination, which indicates a similar trend to the one observed for the non-diffractive fraction (see Fig.~\ref{fig:fnd}). The statistical uncertainty on the \sigmaeff\ measurements ranges from 0.02 to 0.07~mb.

\begin{figure}
	\centering
	\includegraphics[scale=0.47]{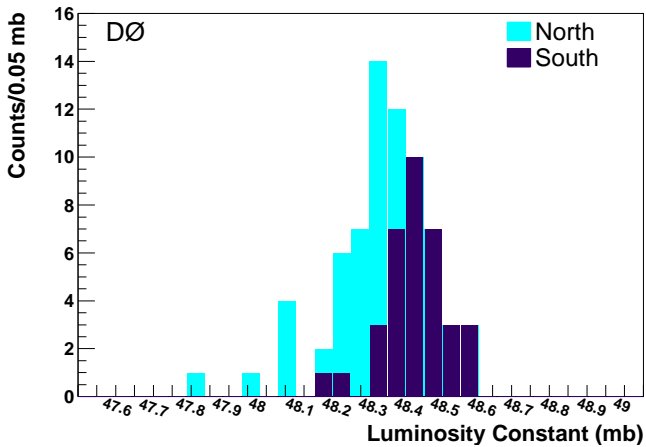}
	\caption {Distribution of luminosity constant measurements.}
	\label{fig:cst_dist}
\end{figure}

The distribution of the \sigmaeff\ measurements is shown in Fig.~\ref{fig:cst_dist}. The average of the north and south luminosity constants gives 
\begin{equation}
	\sigma_{LM} = 48.3~\textrm{mb}.
\end{equation}
The north and south single sided cross sections are determined to be
\begin{equation}
	\sigma_N = \sigma_S = 4.5~\textrm{mb}.
\end{equation}

	\subsection{\label{subsec:uncertaintyt}Luminosity Constant Uncertainty}
		In evaluating the systematic uncertainties affecting the measurement of \sigmaeff, the different sources are propagated through the analysis chain to establish the effect on \sigmaeff, including re-calculating the LM acceptances and determining the non-diffractive fraction for each data sample. This allows us to take into account the correlation between the LM acceptances and the non-diffractive fraction. The different sources of systematic uncertainties are listed below:

\begin{description}

	\item[Inelastic Cross Section] \hfill \\
		\hfill \\
		The CDF and \DZero\ experiments have adopted~\cite{sigma_inel} an inelastic cross section of $\sigma_{inel} = 60.7 \pm 2.4$~mb at 
		$\sqrt{s} = 1.96$~TeV for their luminosity measurements. Propagating the 2.4~mb uncertainty in the inelastic cross section gives an 
		uncertainty of 1.91~mb on the luminosity constant, \sigmaeff.
	\item[Single-Diffractive Fraction]  \hfill \\
		\hfill \\
		The single-diffractive fraction is taken to be $f_{sd} = 0.57 \pm 0.21$~\cite{f_sd}, corresponding to a large
		variation in the single-diffractive cross section. This corresponds to an uncertainty of 0.43~mb on \sigmaeff.
	\item[Time Variation / Radiation Damage] \hfill \\
		\hfill \\
		Periodic adjustments to the PMT high voltage are performed, and the LM scintillators are replaced during long shutdowns to minimize 
		the impact of radiation damage. The high voltage changes typically lead to less than 0.5\% change in the measured luminosity. An 
		uncertainty of $\pm0.5\%$ (0.24~mb) is assigned to \sigmaeff\ due to time variation in the luminosity measurement.
	\item[\textsc{GEANT} Energy Cutoffs] \hfill \\
		\hfill \\
		The simulated events are reprocessed through the \DZero\ \textsc{geant3}~\cite{geant} simulation with lower energy cutoffs. The 
		$\delta$ ray production cutoff is lowered from 1~MeV to 10~keV, the neutral and charged hadron cutoffs are lowered from 1~MeV 
		to 100~keV, and the muon cutoff is lowered from 10~MeV to 100~keV. The 100~keV energy cutoff 
		selected for this study is based on an estimate of the lowest energy that could cause an LM counter to detect a hit. A change of 
		0.24~mb in \sigmaeff\ is observed, and this is assigned as a systematic uncertainty.
	\item[Monte Carlo Material Model] \hfill \\
		\hfill \\
		Secondary interactions in the beam pipe assembly and parts of the silicon microstrip tracker significantly increase the multiplicity 
		of small angle charged particles detected by the LM. After extensive cataloging and modeling of this material, it was necessary to include
		additional material in front of the LM in order to have the \textsc{geant3} model reproduce the observed hit multiplicity distribution.
		MC events have been generated both with the nominal (untuned) material model and the tuned material model. 
		The resulting change in \sigmaeff\  of 0.16~mb is taken as a systematic uncertainty.
	\item[Luminosity Monitor Acceptance] \hfill \\
		\hfill \\
		The acceptance of the LM shows a small dependence (less than 0.1\%) due to the longitudinal variation in the location of the 
		$p \bar{p}$ interaction vertex. In addition, the variations in the transverse position of the interactions\footnote[2]{The beam position 
		could vary as much as 0.3~mm around the nominal transverse location during normal data taking, and the nominal transverse 
		location was 1.6~mm off center before October 2007 and was moved to 0.2~mm off center for stores after that time.} results in a 
		0.17\% variation in the LM acceptance. 
		Adding these effects in quadrature with the Monte Carlo statistics yields an uncertainty of 0.11~mb on $\sigma_{LM}$.
	\item[Light Collection / Radiation Damage] \hfill \\
		\hfill \\
		Radiation damage can reduce the light collection efficiency~\cite{radiation_damage}. To estimate this contribution, 
		MC events with a piece-wise linear change in the light collection efficiency were generated. 
		The light collection efficiency is reduced by a factor of two at the inner edges of the scintillator wedge, increasing linearly 
		to no change at the center of the PMT, and then decreasing linearly to a factor of two reduction at the outer edge of the wedge. The 
		charge threshold is also adjusted to emulate the effect of the HV changes that work to keep the average charge constant. 
		These changes in the MC efficiency calculation resulted in a change in \sigmaeff\ of 0.09~mb, which is assigned as a systematic
		uncertainty.
	\item[North -- South Asymmetry] \hfill \\
		\hfill \\
		The north and south \sigmaeff\ measurements differ by an average of 0.18~mb. Since the final value of \sigmaeff\ is selected to be 
		the average of the north and south measurements, a systematic uncertainty of 0.09~mb is assigned to \sigmaeff\ to account for the 
		observed north -- south asymmetry.
	\item[Luminosity Dependence] \hfill \\
		\hfill \\
		In the ensemble of 35 \sigmaeff\ measurements made at varying luminosities, an RMS spread of 0.08~mb is observed, so this value is 
		assigned as a systematic to account for the observed luminosity dependence.
	\item[\textsc{PYTHIA} Tune] \hfill \\
		\hfill \\
		Non-diffractive events have been generated with a modified value for the transition point between the low-\pT\ and high-\pT\ models, as an 
		alternative to modifying the material model. This is done by changing the \pythia~\cite{pythia} parameter PARP(82), which is known to 
		have a significant affect on the average multiplicity, from the default value of 2.0 to 1.25. 
		This change in the \pythia\ parameter led to a change in \sigmaeff\ of 0.04~mb. 

		Diffractive processes have also been generated with a modified fragmentation parameter MSTP(101). Varying this parameter over its entire
		range results in a change in $\sigma_{LM}$ of 0.07~mb. 

		Combining these two effects in quadrature leads to a change in $\sigma_{LM}$ of 0.08~mb, which is assigned as a systematic uncertainty.
	\item[PDF Choice] \hfill \\
		\hfill \\
		The standard \DZero\ PDF set is CTEQ6L1~\cite{cteq}. Events were also generated using MRST2004NLO~\cite{mrst}. This change 
		resulted in a difference of 0.06~mb in \sigmaeff, which is assigned as an uncertainty.
	\item[Background Unfolding] \hfill \\
		\hfill \\
		The \sigmaeff\ calculation is repeated using multiplicity histograms acquired during the last bunch crossing of a bunch train.
		These bunches have $\approx$40\% higher background than the first bunch of a bunch train. The 
		difference of 0.03~mb between the two calculations is assigned as a systematic uncertainty associated with the background 
		unfolding.
	\item[\textsc{GEANT} Hadronic Model] \hfill \\
		\hfill \\
		For this study, the GCalor~\cite{gcalor} hadronic model is replaced with the Geisha~\cite{geisha} model due to its ability to better 
		handle low energy particle interactions. This change in the \textsc{geant} hadronic model resulted in a 0.03~mb change in \sigmaeff, 
		which is assigned as a systematic uncertainty. 
	\item[Seasonal Timing Variation] \hfill \\
		\hfill \\
		There are seasonal drifts in the \DZero\ clock stemming from temperature variations that result in expansion or contraction 
		of the long cable used to send signals from the accelerator control room to indicate collisions. To account for the effect of seasonal timing 
		observed, the timing window of $\pm6.4$~ns that defines 
		an in-time hit is shifted by $\pm$1~ns. The resulting change of 0.02~mb in \sigmaeff\ is assigned as a systematic uncertainty.
	\item[Charge Threshold] \hfill \\
		\hfill \\
		The charge threshold in the MC simulation is shifted by $\pm$2~pC around the nominal value of 8~pC to account for uncertainties
		in the modeling of the charge threshold. The resulting change of 0.01~mb in \sigmaeff\ is assigned as a systematic uncertainty.

\end{description}

The luminosity constant (\sigmaeff) uncertainties are summarized in Table~\ref{tab:uncertainties}.
\begin{table}
\caption{Contributions to the luminosity constant (\sigmaeff) uncertainty.}
	\begin{ruledtabular}
		\begin{tabular}{lcc}
                  \multicolumn{2}{c}{\textbf{Source}}     & \textbf{Uncertainty (mb)} \\ \hline
                  Inelastic Cross Section		 & &   $\pm$1.91   \\     
                  Single-Diffractive Fraction		 & &   $\pm$0.43   \\
                  Time Variation / Radiation Damage	 & &   $\pm$0.24   \\
                  \textsc{GEANT} Energy Cutoffs		 & &   $\pm$0.24   \\
                  Monte Carlo Material Model		 & &   $\pm$0.16   \\                  
                  Luminosity Monitor Acceptance		 & &   $\pm$0.11   \\  
                  Light Collection / Radiation Damage	 & &   $\pm$0.09   \\
                  North -- South Asymmetry		 & &   $\pm$0.09   \\
                  Luminosity Dependence			 & &   $\pm$0.08   \\   
                  Pythia Tune				 & &   $\pm$0.08   \\
                  PDF Choice				 & &   $\pm$0.06   \\
                  Background Unfolding			 & &   $\pm$0.03   \\
                  \textsc{GEANT} Hadronic Model	         & &   $\pm$0.03   \\
                  Seasonal Timing Variation		 & &   $\pm$0.02   \\
                  Charge Threshold                       & &   $\pm$0.01   \\

		\end{tabular}
	\end{ruledtabular}
\label{tab:uncertainties}
\end{table}
Adding the uncertainties in quadrature yields an uncertainty in \sigmaeff\ of \uncertainty, where 1.9~mb is associated with the uncertainty in the inelastic cross section, 0.4~mb is associated with the uncertainty in the single-diffractive fraction, and 0.4~mb is associated with the remaining sources of uncertainty.
\hfill \\

\section{\label{sec:integrated}Integrated Luminosity}
		The integrated luminosity \IntLumi, defined as
\begin{equation}
	{\cal{L}} = \int_0^T L \cdot dt,
\end{equation}
where $T$ is the data taking period, is the relevant quantity used in the measurements of cross sections and in setting upper limits on the production of new particles. In addition to the uncertainty on the luminosity constant, the uncertainty on the determination of the integrated luminosity takes into account additional contributions that cover possible variations with time and with luminosity of the luminosity constant. These additional sources of uncertainty are discussed below.

The ``delivered'' luminosity is the integrated luminosity delivered by the Tevatron. The ``recorded'' luminosity is the integrated luminosity associated with a specific trigger and takes into account the deadtime and losses in the data acquisition system. Level 1 triggers are grouped together so that they have common deadtime, i.e., common sources of enable, disable, and readout~\cite{d0}. The recorded integrated luminosity referred to in this section corresponds to the luminosity exposure of the experiment's  jet trigger with the highest transverse energy, $E_T = E \cdot \mathrm{sin} \theta$, which requires at least one jet with $E_T > 125$~GeV.

The total recorded integrated luminosity for Run IIb is assessed to be \integrated~fb$^{-1}$. Of that, $\sim$0.2~fb$^{-1}$ were recorded with luminosity above 300~$\mu \mathrm{b^{-1} s^{-1}}$, and $\sim$1.7~fb$^{-1}$ were recorded with luminosity between 200 and 300~$\mu \mathrm{b^{-1} s^{-1}}$. 
\begin{figure}
	\centering		
	\includegraphics[scale=0.47]{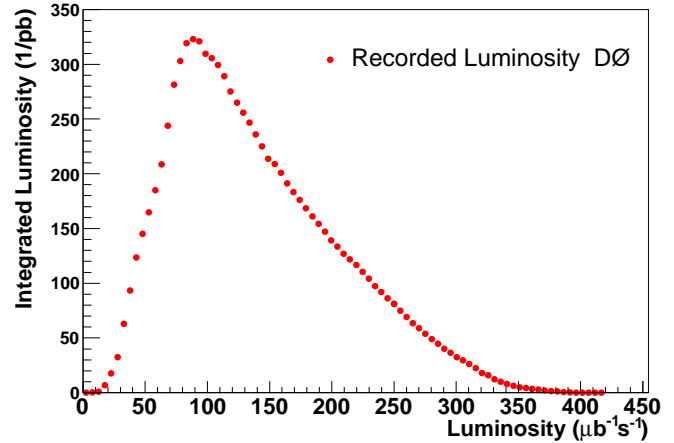}
	\caption {The recorded luminosity profile in bins of 5~$\mu \mathrm{b^{-1} s^{-1}}$.}
	\label{fig:profiles}
\end{figure}
Figure~\ref{fig:profiles} shows the recorded luminosity profile, in bins of width 5~$\mu \mathrm{b^{-1} s^{-1}}$, for the full Run IIb dataset.
A nominal 60~s measurement period is used for the delivered luminosity measurement, which is sufficient to provide an accurate luminosity measurement up to the highest luminosities in the Run IIb dataset (see Appendix~\ref{app:empty}).

\subsection{\label{subsec:LumiStability}Long-Term Stability of the Luminosity Measurement} 

The yield of single muons in the forward muon system~\cite{d0muon} can be used as an independent check of the stability of the luminosity measurements. The single muon yield $Y$ is monitored regularly using special data samples and is obtained by 
\begin{equation}
	Y = \frac{N_\mu}{{\cal{L}}},
	\label{eq:muonYields}
\end{equation}
where $N_\mu$ is the number of the forward muons, and \IntLumi\ is the integrated luminosity of the respective data sample.  
\begin{figure}
	\centering		
	\subfigure{\label{fig:muonYieldsTime}        \includegraphics[scale=0.47]{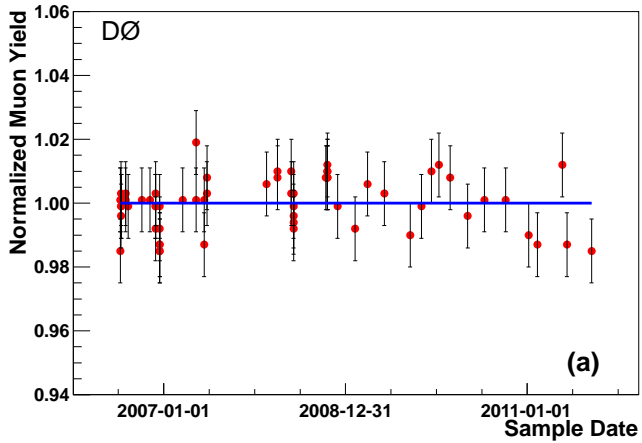}}
	\subfigure{\label{fig:muonYieldsTimeProj} \includegraphics[scale=0.45]{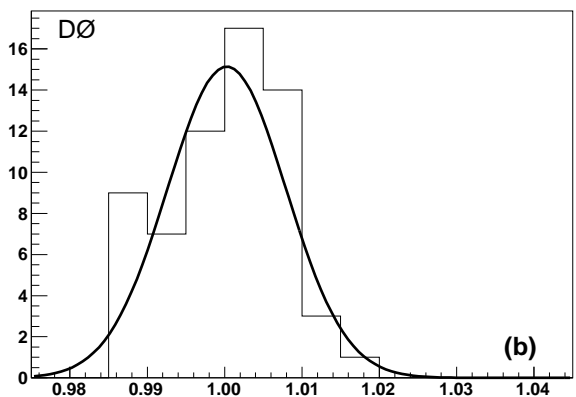}}	
	\caption {Normalized single muon yields (a) as a function of time, and (b) its projection on the $y$-axis.}
	\label{fig:muonYields_time}
\end{figure}
The normalized yield is obtained by dividing the value of the muon yield for each data sample by the mean value of all the data samples collected
\begin{equation}
	Y_\mathrm{norm} = \frac{Y}{\mathrm{mean~value}}.
\end{equation}
Figure~\ref{fig:muonYieldsTime} shows the normalized yield measurements for Run IIb. If the statistical error on $N_\mu$ is less than 1\% the uncertainty on the yield, $\sigma_Y$, is set to 1\%; otherwise $\sigma_Y$ is set to the statistical error of $N_\mu$. Figure~\ref{fig:muonYieldsTimeProj} shows the distribution of the normalized muon yields super-imposed with a Gaussian function that indicates a typical variation of $\sim$0.8\%.

\begin{figure}
	\centering		
	\subfigure{\label{fig:muonYieldsLumi}        \includegraphics[scale=0.47]{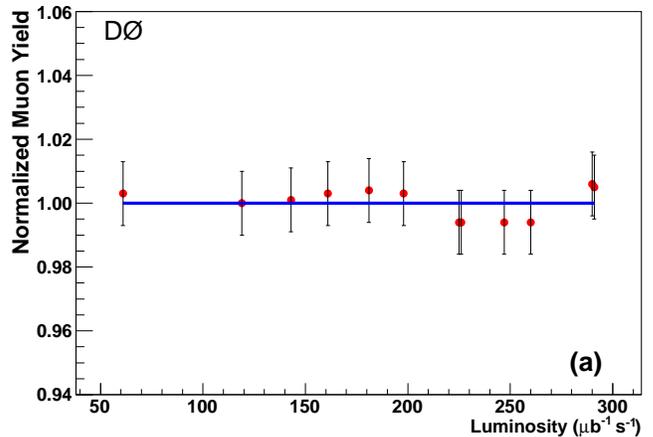}}
	\subfigure{\label{fig:muonYieldsLumiProj} \includegraphics[scale=0.45]{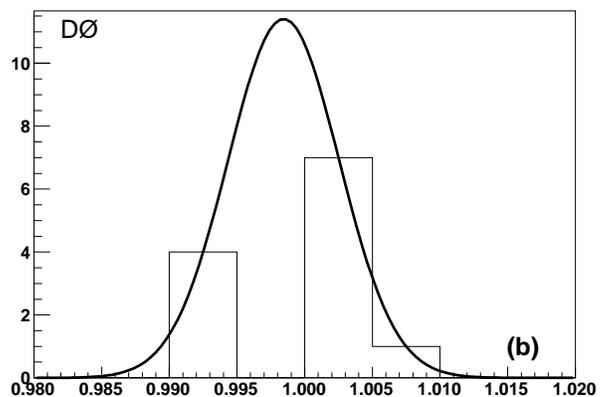}}	
	\caption {Normalized single muon yields (a) as a function of luminosity, and (b) its projection on the $y$-axis.}
	\label{fig:muonYields_lumi}
\end{figure}

In addition, muon yields are measured as a function of luminosity. Figure~\ref{fig:muonYields_lumi}
shows an example of the distribution of the normalized muon yields collected during two Tevatron stores. The super-imposed Gaussian on Fig.~\ref{fig:muonYieldsLumiProj} indicates a typical variation of $\sim$0.4\%. After accounting for the statistical errors on $N_\mu$ and the uncertainty due to the time variation / radiation damage already included in the luminosity constant, a systematic uncertainty of 0.6\% is assigned to the integrated luminosity measurement based on the observed variation in the muon yield.

 \subsection{\label{subsec:IntLumiUncertainty}Integrated Luminosity Measurement Uncertainty}
 
 The uncertainty of the integrated luminosity includes a dominant contribution of 4.2\% stemming from the uncertainty on the luminosity constant. Additionally, it includes an uncorrelated contribution of 0.6\% in the integrated luminosity measurement described in the previous section.
 Other potential source of systematic uncertainty includes events that are lost in the \DZero\ data acquisition and reconstruction system and are not properly taken into account. The fraction of these events has been studied and found to be negligible ($ << 0.1\%$).
 
 Adding these uncertainties in quadrature yields an uncertainty in the integrated luminosity of 4.3\%, where 4.0\% originates from the uncertainty in the inelastic cross section and is correlated with both the CDF and \DZero\ Run IIa integrated luminosity measurements, 0.9\% originates from the uncertainty in the single diffractive fraction and is uncorrelated with the CDF integrated luminosity measurement, but correlated with the \DZero\ Run IIa integrated luminosity measurement, and 1.1\% originates from the remaining sources of uncertainty that are uncorrelated with both the CDF and \DZero\ Run IIa integrated luminosity measurements.

\section{\label{sec:summary}Summary and Conclusions}
	In summary, we have measured the effective inelastic cross section, \sigmaeff, as seen by the \DZero\ luminosity monitor, and assessed the recorded integrated luminosity for data collected with the \DZero\ detector at the Fermilab Tevatron Collider for the period called Run IIb (June 2006 to September 2011). A luminosity constant of $\sigma_{LM} = 48.3 \pm 2.0$~mb is obtained. The Run IIb luminosity constant is 0.6\% larger than the Run IIa luminosity constant ($48.0 \pm 2.9$~mb) and its uncertainty has been reduced from 6.1\% to 4.2\%. 

The recorded integrated luminosity for the highest $E_T$ jet trigger is \IntLumi\ = \integrated\ $\pm$ \intUncertainty\ during Run IIb. 
The total relative uncertainty of the Run IIb \DZero\ recorded integrated luminosity is determined to be 4.3\%, where 4.0\% is associated with the inelastic cross section, 0.9\% is associated with the single diffractive fraction, and 1.1\% is associated with the \DZero\ sources of uncertainty.

%
We thank the staffs at Fermilab and collaborating institutions, and all our \DZero\ collaborators for their support. In particular we would like to extend our gratitude to Brad Abbott of the University of Oklahoma and David Hedin of the Northern Illinois University for the thorough review of this document and their insightful comments, Margherita Vittone -- Wiersma of the Fermilab Computing Division for her tireless work on the luminosity database, and Andrey Shchukin of the Institute for High Energy Physics, Protvino, Russia for providing the muon yields data and plots.  
We also acknowledge support from the
DOE and NSF (USA);
CEA and CNRS/IN2P3 (France);
FASI, Rosatom and RFBR (Russia);
CNPq, FAPERJ, FAPESP and FUNDUNESP (Brazil);
DAE and DST (India);
Colciencias (Colombia);
CONACyT (Mexico);
KRF and KOSEF (Korea);
CONICET and UBACyT (Argentina);
FOM (The Netherlands);
STFC and the Royal Society (United Kingdom);
MSMT and GACR (Czech Republic);
CRC Program and NSERC (Canada);
BMBF and DFG (Germany);
SFI (Ireland);
The Swedish Research Council (Sweden);
and
CAS and CNSF (China).

\appendix

\section{Empty Crossing Method}
	\label{app:empty}
		The \DZero\ luminosity is derived from the number of beam crossings with north -- south in-time coincidences in the \DZero\ LM that occur during a measurement period. The luminosity reported to the Fermilab accelerator division for monitoring purposes is based on a nominal 15~s measurement period, while the luminosity used for physics analyses employs a nominal 60~s measurement period. The actual measurement period is occasionally shorter to ensure synchronization between the luminosity measurement and the state of the data acquisition system. Statistical fluctuations in the number of luminosity coincidences lead to statistical and systematic errors in individual luminosity measurements.

The luminosity calculation is performed separately for each of the 36 beam bunches in the Tevatron, resulting in each bunch having its own measured luminosity. For a bunch with true luminosity $L$, the average number of proton -- antiproton interactions that produce north -- south coincidences is proportional to the luminosity
\begin{equation}
	\overline{N}_{NS} ( L ) = \frac{\sigma_{LM}}{f} L, 
\end{equation}
where \sigmaeff\ is the effective inelastic cross section seen by the LM and $f$ is the beam crossing frequency. Similarly, the average number of interactions in a beam crossing that produce in-time hits in only the north (south) LM array is given by
\begin{equation}
	\begin{split}
		\overline{N}_N (L) & = \frac{\sigma_N}{f} L, \\
		\overline{N}_S (L) & = \frac{\sigma_S}{f} L,  \\
	\end{split}
\end{equation}
where $\sigma_N$ and $\sigma_S$ are the single-sided effective cross sections.

The fraction of beam crossings that do not produce a north-south coincidence of in-time LM hits and are classified as empty is given by Poisson statistics 
\begin{equation}
	\begin{split}
		 F_0 (L) & = e^{-\overline{N}_{LM}(L)} \\
			& \cdot  \left ( e^{-\overline{N}_{N}(L)} + e^{-\overline{N}_{S}(L)} - e^{- ( \overline{N}_{N}(L) + \overline{N}_{S}(L) )} \right ),
	\end{split}
\end{equation}
where the first factor is the probability for having no $p \bar{p}$ interactions giving a north-south coincidence. The term in the parenthesis gives the probability for not having multiple single-sided interactions that result in a north-south coincidence.

The average number of empty crossings during a measurement period is then
\begin{equation}
	\overline{N}_0 (L) = N_\mathrm{Live} \times F_0 (L),
\end{equation}
where $N_\mathrm{Live}$ is the number of live beam crossings during the measurement period. In the results shown below, we ignore the small fraction of beam crossings that are rejected due to halo veto ( $\sim$1\%) and we consider $N_\mathrm{Live} = f \cdot T$, where $T$ is the measurement period.

\subsection{\label{subsec:HighLumiPerformance} Behavior of the \DZero\ Luminosity Measurement at High Luminosity} 

The number of empty crossings observed in different measurement periods can be described with the binomial distribution. In the high luminosity limit where the number of empty crossings is small, the distribution of the observed number of empty crossings is well approximated by a Poisson distribution
\begin{equation}
	P(n_0) = \frac{n^{\overline{N}_0 (L)}}{n!} e^{-\overline{N}_0(L)},
\end{equation}
where $P(n_0)$ is the probability of observing $n_0$ empty crossings.

For each measurement period, the measured luminosity $L_m$ corresponding to the observed number of empty beam crossings, $n_0$, is calculated by numerically solving the equation
\begin{equation}
	\begin{split}
	 n_0 & = N_{Live} e^{-\sigma_{LM} L_m / f} \\
	& \cdot \left ( e^{-\sigma_N L_m / f} + e^{-\sigma_S L_m / f} - e^{-(\sigma_N + \sigma_S) L_m / f} \right ) .
	\end{split}
\end{equation}
In the case where no empty crossings are observed, the solution of the above equation yields an infinite measured luminosity. When that occurs, the luminosity is set to the value that would be found if there had been one empty crossing observed. The impact of this approximation on the measurement of the luminosity in \DZero\ is discussed below.

The average measured luminosity is given by
\begin{equation}
	\overline{L}_m = P(0) L_m (1) + \sum_{\substack{n_0 = 1}}^{\substack{fT}} P(n_0) L_m (n_0)
\end{equation}
where $P(n_0)$ is the probability of observing $n_0$ empty crossings, and the first term accounts for the special handling where no empty crossings are observed.

The total luminosity is obtained by summing the luminosity from the 36 beam bunches. If all 36 bunches had the same luminosity, the total luminosity would be 36 times the bunch luminosity and the RMS spread of the total luminosity would be a factor of six times the RMS for a single bunch since each bunch measurement is statistically independent. In practice there are typically a few percent variations among the bunch luminosities. While these bunch-to-bunch variations are accounted for in the \DZero\ luminosity measurement, the results following illustrate the behavior of the measured total luminosity under the assumption that all bunches have the same bunch luminosity. Including typical bunch-to-bunch variations will not significantly affect these results.

\begin{figure}
	\centering
	\includegraphics[scale=0.47]{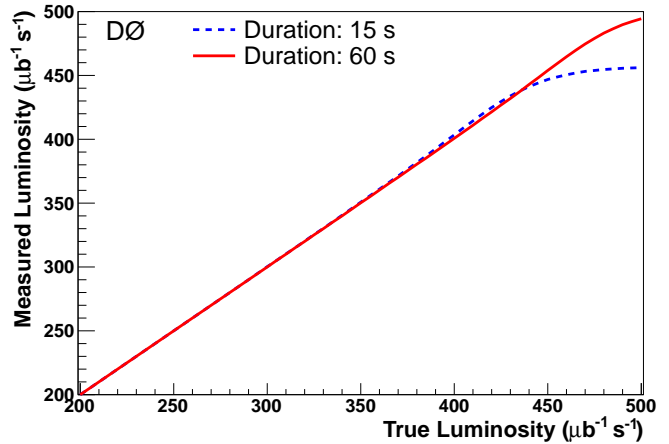}
	\caption {Average measured luminosity versus the true luminosity for 15~s and 60~s measurement periods.}
	\label{fig:relation}
\end{figure}

The relation between the average measured luminosity and the true luminosity for measurements periods of 15~s and 60~s is shown in Fig.~\ref{fig:relation}. Due to the non-linear behavior in the empty crossing probability at high luminosity, the average measured luminosity systematically exceeds the true luminosity before entering the saturation region, where the luminosity asymptotically approaches the value $L_m(1)$ that is assigned to bunches with less than two empty beam crossings. 

Table~\ref{tab:non-linearity} illustrates how the non-linear behavior of the empty crossing probability and the special treatment of the case where there are no empty crossings in the measurement period lead to non-linear behavior for the average measured luminosity. In this example, the true luminosity is 420~$\mathrm{\mu b^{-1} s^{-1}}$, the measurement period is 15~s, and the average number of empty crossings in the measurement period is 3.05. The exponential decrease in the measured luminosity as the number of observed empty crossings increases results in an average measured luminosity of 424.8~$\mathrm{\mu b^{-1} s^{-1}}$, which is 1.1\% higher than the true luminosity.
\begin{table}
\caption{Non-linearity in the luminosity measurement showing the probability, measured luminosity, and contribution to the average luminosity as a function of the observed number of empty crossings. In this example, the true luminosity is 420~$\mathrm{\mu b^{-1} s^{-1}}$ and the measurement period is 15~s.}
	\begin{ruledtabular}
		\begin{tabular}{c|ccc}
			$n_0$ & $P(n_0)$ & $L_m$ ($\mathrm{\mu b^{-1} s^{-1}}$)  & $P(n_0) \cdot L_m$ ($\mathrm{\mu b^{-1} s^{-1}}$)  \\  \hline
			 0	   &       0.047  &                       456.9                                     &                                 21.7          					 \\
			 1	   &       0.145  &                       456.9                                     &                                 66.1          					 \\
			 2	   &       0.220  &                       434.0                                     &                                 95.7          					 \\
 			 3	   &       0.224  &                       420.5                                     &                                 94.2          					 \\			 			 
			 4	   &       0.171  &                       411.0                                     &                                 70.1          					 \\
			 5	   &       0.104  &                       403.6                                     &                                 42.0          					 \\
			 6	   &       0.053  &                       397.5                                     &                                 21.0          					 \\
			 $> 6$ &       0.036  &                       --	                                          &                                 14.0          					 \\ \hline
			 Sum	   &       1	      &                       --                                             &                                 424.8          					 \\			 			 			 
		\end{tabular}
	\end{ruledtabular}
\label{tab:non-linearity}
\end{table}

\begin{figure}
	\centering
	\includegraphics[scale=0.47]{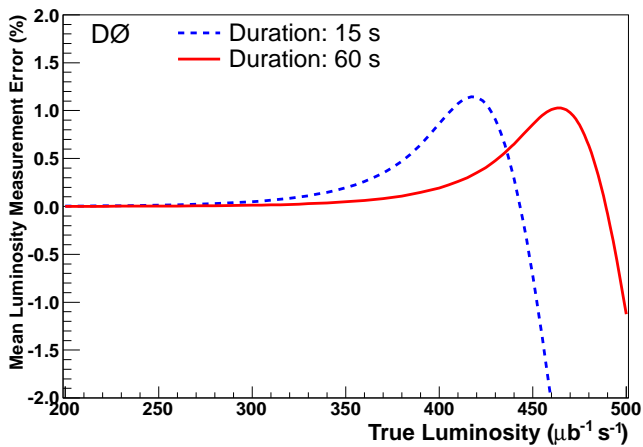}
	\caption {Mean deviation between the average measured luminosity and the true luminosity for 15~s and 60~s measurement periods.}
	\label{fig:deviation}
\end{figure}

Figure~\ref{fig:deviation} shows the mean deviation between the average measured luminosity and the true luminosity as a function of the true luminosity. The largest positive deviation occurs when the average number of empty crossings is $\sim 3$. This occurs at a luminosity of 420~$\mathrm{\mu b^{-1} s^{-1}}$ for a 15~s measurement period, where the average number of interactions with a north-south coincidence is $\sim 12$ and the fraction of empty crossings is $\sim 4 \cdot  10^{-6}$. For a 60~s measurement period, an average of 3 empty crossings occurs at a luminosity of 470~$\mathrm{\mu b^{-1} s^{-1}}$, where the average number of interactions with a north-south coincidence is $\sim 13$ and the fraction of empty crossings is $\sim 10^{-6}$. Given that only a small fraction of data was collected at luminosities in excess of 300~$\mathrm{\mu b^{-1} s^{-1}}$ and that the measurement period used for the determination of the integrated luminosity is of 60~s, we estimate that the impact of the non-linear behavior in the empty crossing probabilities has a negligible impact on the precision of the integrated luminosity determination in \DZero.

\begin{figure}
	\centering
	\includegraphics[scale=0.47]{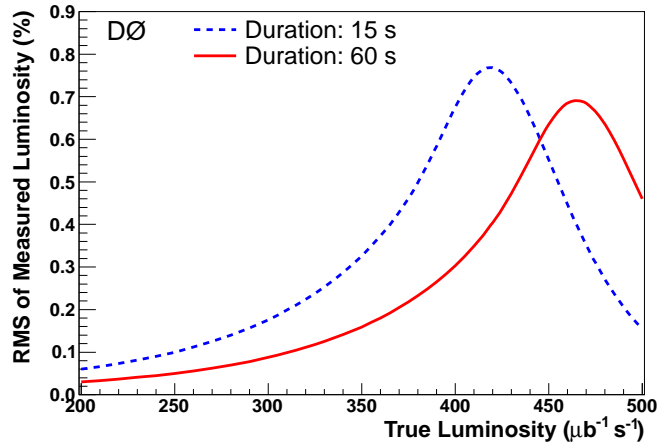}
	\caption {RMS width of the measured luminosity for 15~s and 60~s measurement periods.}
	\label{fig:rms}
\end{figure}

The RMS width of the measured luminosity for a single bunch is given by
\begin{equation}
	\sigma^2_{L_m}  = P(0) \left ( L_m(1) - \overline{L}_m \right )^2 + \sum_{\substack{n_0 = 1}}^{\substack{fT}} P(n_0) \left ( L_m(n_0) - \overline{L}_m \right )^2,
\end{equation}
where the first term accounts for the special handling where there are no empty crossings observed. The statistical uncertainty in the total luminosity for the 36 bunches is shown in Fig.~\ref{fig:rms}. The statistical uncertainty in the luminosity measurement is less than 0.1\% for luminosities in the range 1.5 -- 250~$\mathrm{\mu b^{-1} s^{-1}}$ for the 15~s measurement period and 0.4 -- 310~$\mathrm{\mu b^{-1} s^{-1}}$ for the 60~s measurement period. 
The RMS width reaches a maximum of $\sim 0.7$\% when the average number of empty crossings is $\sim 3$. Further increases in the true luminosity push the luminosity measurement into the saturation region where an increasing fraction of luminosity measurements report the maximum possible measured luminosity $L_m(1)$, leading to a decrease in the RMS width.

At high luminosity, the statistical error can be approximated by
\begin{equation}
	\frac{\sigma_L}{L} \approx \frac{1}{ \sqrt{N_B} } \frac{1}{ \overline{N}_{NS} } \frac{\sigma_{R_0}}{ R_0 },
\end{equation}
where $R_0$ is the rate of empty crossings, $N_B$ is the number of beam bunches, and the small contribution to the empty crossing rate from multiple single-sided interactions is ignored. Thus, the large statistical error on the empty crossing rate when there are an average of 3 empty crossings ($1/\sqrt{3}$ or 58\%) is reduced by a factor of six due to the 36 independent bunch measurements and an additional factor of $\sim 12$ at 420~$\mathrm{\mu b^{-1} s^{-1}}$ due to the exponentially falling empty crossing rate to yield a precision in the total luminosity of better than 1\%.

A key requirement for the empty crossing method to work at high luminosity is that beam crossings with in-time north-south coincidences are not misclassified as empty crossings. With empty crossings rates in the part-per-million range at peak luminosities, this misclassification probability must be well under $10^{-6}$ for typical beam crossings. Such beam crossings are easily identified in the \DZero\ LM since a typical beam crossing yields in-time hits in most or all of the 24 luminosity counters, whereas the requirement is $\geq 1$ in-time hit. Since early hits in a large number of counters can mask the presence of in-time hits, a halo veto has been implemented to exclude beam crossings with more than six early hits from the luminosity calculation. Timing distributions are monitored and the LM TDCs are recalibrated as needed to ensure that the timing distributions are well centered within the timing window. Tests of the digital counting electronics have proven this system to be robust with no evidence of misclassification.

\section{Background Studies}
	\label{app:background}
		\subsection{Luminosity Dependence of Background Rate}

\begin{figure}[b]
	\centering
	\subfigure{\label{fig:appB1} \includegraphics[scale=0.43]{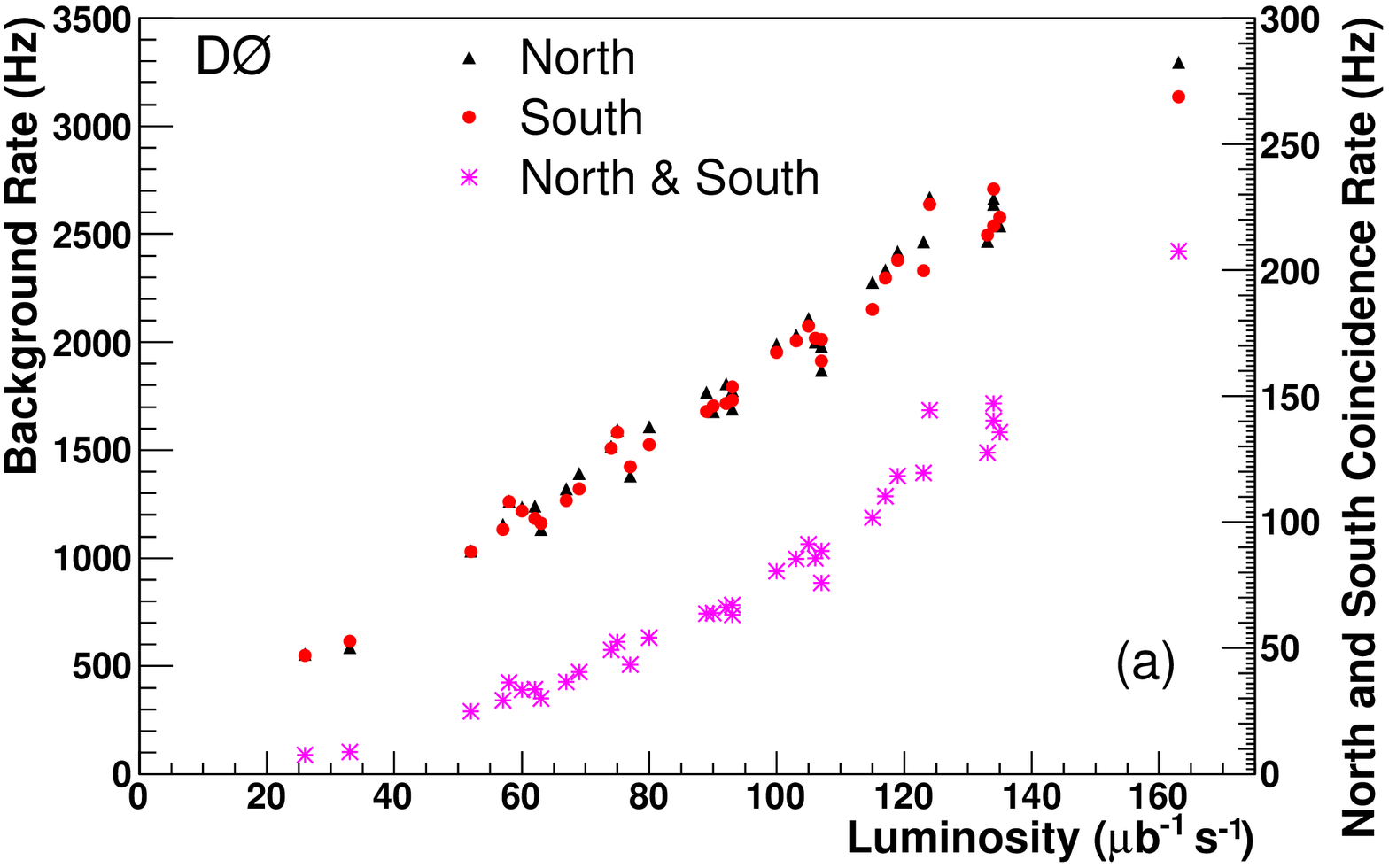}}
	\subfigure{\label{fig:appB2} \includegraphics[scale=0.43]{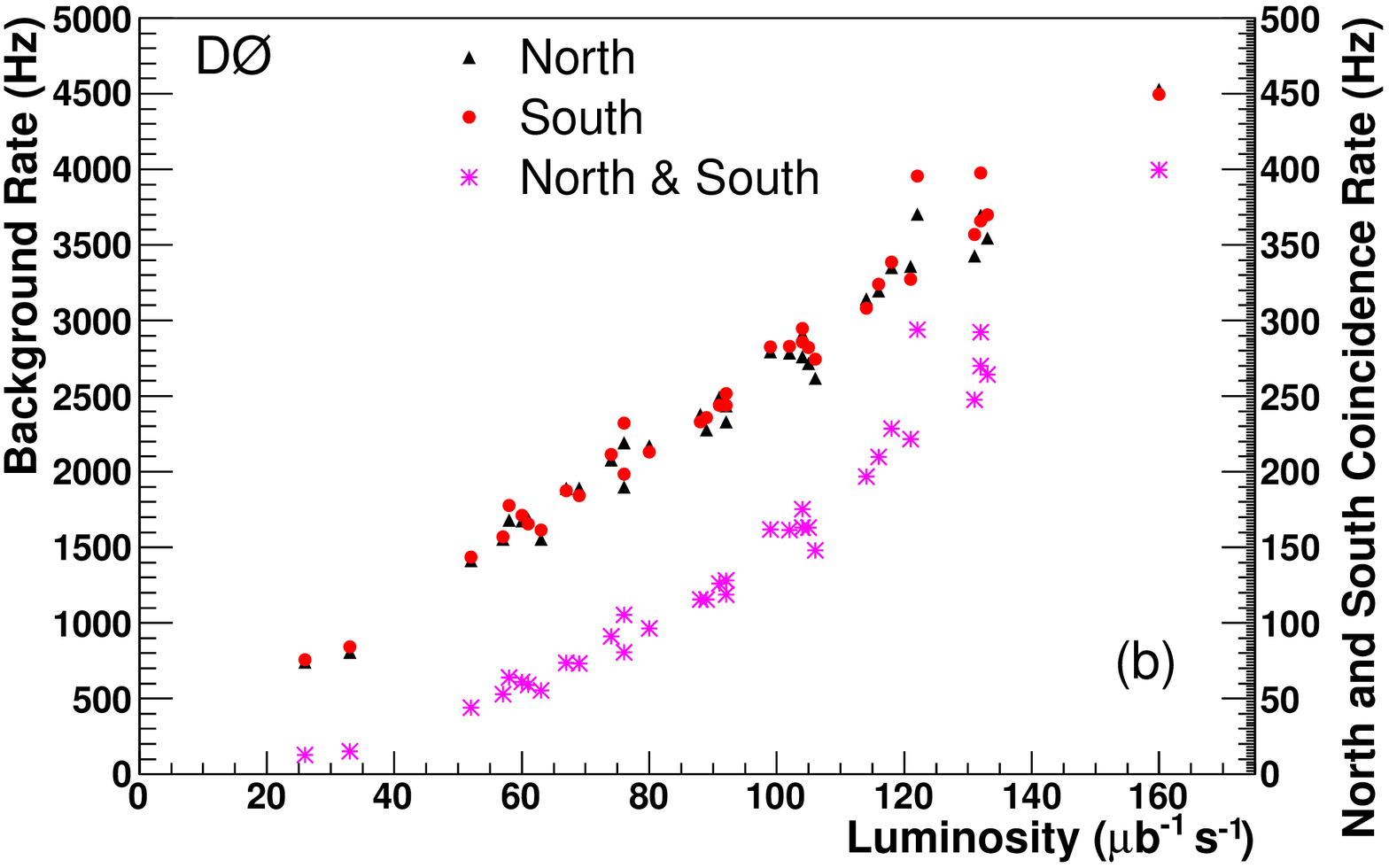}}
	\caption {Background rate, measured in an empty tick (a) immediately before the first bunch crossing and (b) 396~ns after the last bunch crossing of a bunch train. One or more background hits in the north/south luminosity monitors is required (left axis). Also shown is the rate when both north and south luminosity monitors have one or more background hits (right axis).}
	\label{fig:appB1and2}
\end{figure}

The luminosity dependence of the rate for background hits in ticks that do not have proton-antiproton collisions is studied, where the 
multiplicity of background hits is neglected and only the fraction of ticks with $ > 0$ hits in the LM is measured. The north and south background rates, and the coincidence rate where background hits occur in both north and south counters are separately measured. Figure~\ref{fig:appB1} shows the background rates as a function of luminosity, measured in an empty tick 132~ns before the first bunch crossing of a bunch train. 
Figure~\ref{fig:appB2} shows the same rates measured in an empty tick 396~ns after the last bunch crossing of a bunch train. In both cases, the north and the south rates scale approximately linearly with luminosity, while the north -- south coincidence rate is much lower and has a non-linear dependence on luminosity.

\subsection{Effective Background Cross Section}
	\label{subsection:effective_background}
Since the north and south background rates scale linearly with the luminosity, the background rate can be treated as an effective background cross section. Since there is no actual luminosity in the ticks where the backgrounds are measured, 1/36 of the \DZero\ luminosity (i.e., the average luminosity attributable to one of the 36 beam bunches) is taken to calculate the background cross section.

To account for pileup in the background, Poisson statistics is used to relate the background cross section and the background rate $R_{BG}$ in a given detector using
\begin{align}
	P(0) & = e^{-\sigma_{BG} L / f} = 1 - \frac{R_{BG}}{f} \\
	\sigma_{BG} & = -\frac{f}{L} \ln \left ( 1 - \frac{R_{BG}}{f} \right ), 
\end{align}
where $\sigma_{BG}$ is the background cross section, $L$ is the luminosity, $f$ is the beam crossing frequency, and $R_{BG}$ is the background rate.

\begin{figure}
	\centering
	\includegraphics[scale=0.47]{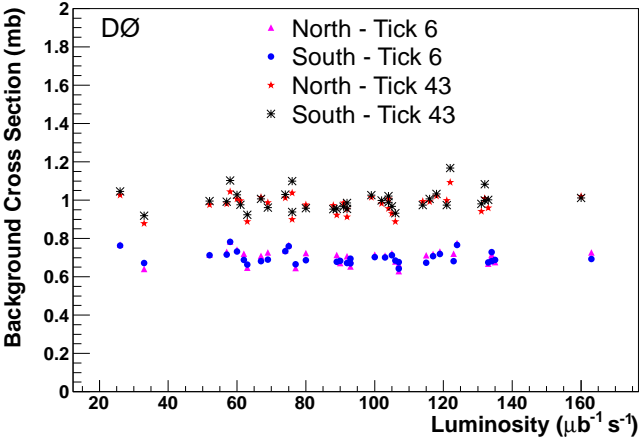}
	\caption {Effective background cross section as a function of luminosity, measured in an empty tick 132~ns before (tick 6) and 396~ns after (tick 43) a bunch train.}
	\label{fig:appB3}
\end{figure}

Figure~\ref{fig:appB3} shows that the background cross section is largely independent of the luminosity. The background cross 
section after a bunch train is about 40\% higher than immediately before the bunch train, indicating that the background cross section increases during the bunch train.

\begin{figure}
	\centering
	\includegraphics[scale=0.47]{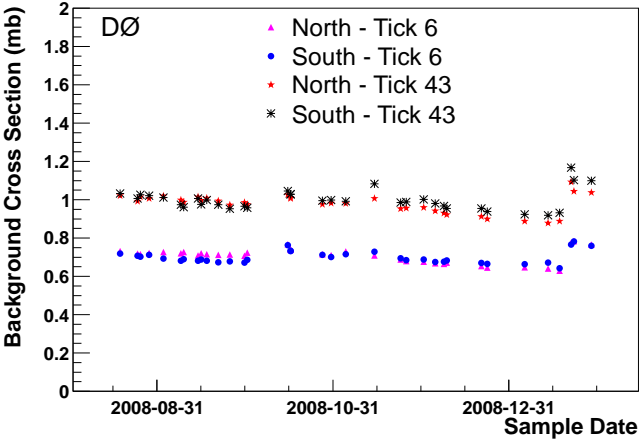}
	\caption {Time dependence of the background cross sections. The increase in the cross section in late January 2009 corresponds to the raising of the PMT high voltages to compensate for radiation damage. The PMT high voltages were adjusted a few days before the first data point in the plot. Tick 6 refers to an empty tick immediately before a bunch train, and tick 43 refers to an empty tick after a bunch train.}
	\label{fig:appB4}
\end{figure}

For a given tick, the background cross section is observed to decrease over time, as shown in Fig.~\ref{fig:appB4},
and attributed to radiation damage to the scintillator. The cross section increases significantly in data recorded in late January 2009 immediately following an increase in the high voltage applied to the LM PMTs to compensate for the radiation damage. There is also an increase in the background rate in early October 2008, following a long downtime of the accelerator, interpreted as evidence of some annealing of the radiation damage in the scintillator following an extended shutdown. Averaging the 140 background cross section measurements from 35 data samples shown in Fig.~\ref{fig:appB3}, the average background cross section is estimated to be: $\sigma_{BG} = 0.85$~mb.

\subsection{Background Characteristics}

The coincidence background where both north and south luminosity monitors recorded hits is studied, since such coincidences are counted in the luminosity measurement. Table~\ref{tab:appB1} shows a high luminosity background multiplicity distribution for an empty tick immediately before a bunch train. 
\begin{table}
\caption{Background multiplicity distribution measured in an empty tick immediately before a beam crossing. The data were acquired at a luminosity of $272~\mu \mathrm{ b^{-1} s^{-1}}$. The rows indicate the north multiplicity and the columns the south multiplicity.}
	\begin{ruledtabular}
		\begin{tabular}{c|ccccc}
			N/S & 0               & 1            & 2           & 3          & $\geq 4 $ \\  \hline
			0     & 4522672 & 466907 & 76071 & 10313 & 3203         \\
			1     & 501848   & 52260    & 8699   & 1265   & 394            \\
			2     & 84467     & 9072      & 1524    & 208     & 63              \\
			3     & 11706     & 1349      & 263      & 33        & 17              \\
			$\geq 4$ & 3460 & 399     & 78        & 23        & 10              \\
		\end{tabular}
	\end{ruledtabular}
\label{tab:appB1}
\end{table}
One or more background hits are present in 21\% of the ticks, and 1.3\% have hits in both north and south luminosity monitors. The multiplicity of background hits is low. Approximately 79\% of the ticks with background hits have only one background hit among the 48 luminosity counters.

The question whether the north and south background hits are statistically independent is studied by projecting the 2D distributions to obtain 1D probability distributions for observing a particular number of hits in the north/south monitor. Table~\ref{tab:appB2} shows the 1D probability distributions derived from the 2D multiplicity distribution in Table~\ref{tab:appB1}, 
\begin{table}[h]
\caption{North and south 1D probability distributions derived from the 2D multiplicity distribution in Table~\ref{tab:appB1}.}
	\begin{ruledtabular}
		\begin{tabular}{c|cc}
			Multiplicity &   $P_N$  & $P_S$ \\  \hline
			0     & 0.8824     & 0.8902 \\
			1     & 0.09806   & 0.0921 \\
			2     & 0.01656   & 0.0151 \\
			3     & 0.0023     & 0.0021 \\
			$\geq 4$ & 0.0007 & 0.0006 \\
		\end{tabular}
	\end{ruledtabular}
\label{tab:appB2}
\end{table}
where $P_N(i)$ is the probability of having $i$ counters hit in the north luminosity monitor and $P_S(j)$ is the probability of having $j$ counters hit in the south luminosity monitor.
\begin{figure}[h]
	\centering
	\includegraphics[scale=0.47]{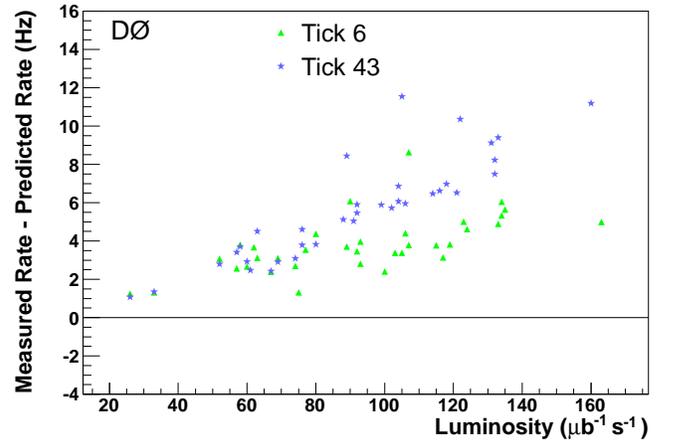}
	\caption {Difference between the measured and predicted rates before (Tick 6) and after (Tick 43) bunch train crossings with hits in both north and south luminosity monitors.}
	\label{fig:appB5}
\end{figure}

If the backgrounds in the north and the south detectors are uncorrelated, the number of entries in the 2D multiplicity distribution, $N(i,j)$, will be given by
\begin{equation}
	N(i,j) = P_N(i)P_S(j)N_0 ,
	\label{eq:n_ij}
\end{equation}
where $N_0$  is the total number of entries in the 2D multiplicity distribution. Table~\ref{tab:appB3} shows the predicted 2D multiplicity distribution using this equation and the 1D probability distributions in Table~\ref{tab:appB2}.
\begin{table}[h]
\caption{Predicted 2D multiplicity distribution obtained from the 1D probability distributions in Table~\ref{tab:appB2} under the assumption that the north and south 1D distributions are statistically independent.}
	\begin{ruledtabular}
		\begin{tabular}{c|ccccc}
			N/S & 0               & 1            & 2           & 3          & $\geq 4 $ \\  \hline
			0     & 4521378 & 467642 & 76444 & 10448 & 3253         \\
			1     & 502477   & 51971    & 8495   & 1161   & 362            \\
			2     & 84865     & 8777      & 1434    & 196     & 61              \\
			3     & 11900     & 1231      & 201      & 27        & 9              \\
			$\geq 4$ & 3534 & 366     & 60        & 8           & 3              \\
		\end{tabular}
	\end{ruledtabular}
\label{tab:appB3}
\end{table}
Reasonably good agreement is observed between the measured 2D multiplicity distributions shown in Table~\ref{tab:appB1} and the predicted distribution in Table~\ref{tab:appB3}. Nevertheless, there is a small but persistent underestimate in the number of predicted coincidences between north and south detectors.

Figure~\ref{fig:appB5} shows the difference between the predicted and measured rate for observing north -- south background coincidences for the data sample considered. 
The predicted rate is consistently underestimated by a small amount, and the magnitude of the discrepancy grows with luminosity.

There are no additional known sources of background that can give north -- south coincidences. In predicting the random coincidence rate, the assumption was made that there are no correlations in the probabilities that the north and south detectors have background hits. Since the rate of background hits scales with luminosity, the background must be associated with beam -- beam interactions in previous beam crossings. The number of interactions in a given beam crossing will fluctuate according to Poisson statistics. Beam crossings with upward fluctuations in the number of interactions will have a higher probability of producing background hits in both the north and south luminosity monitors, while crossings with a downward fluctuation will have a lower probability of producing background hits.

Let $P$ represent the probability of producing a background hit in either the north or south luminosity monitor for a given tick. Instead of assuming a fixed value for $P$, it is assumed that $P$ fluctuates depending on how many beam -- beam interactions took place in recent beam crossings. The probability of having a north -- south coincidence, $P_{coin}$, is given by
\begin{align}
	\left \langle P_{coin} \right \rangle & = \left \langle P^2 \right \rangle \nonumber \\
	& = \left \langle P \right \rangle ^2 + \sigma^2_P , 
\end{align}
where $\sigma_P$ is the RMS spread of the $P$ distribution. If there is an average of $N_{eff}$ beam -- beam interactions producing background hits, we obtain:
\begin{equation}
	\frac{\sigma_P}{ \left \langle P \right \rangle } = \frac{1}{\sqrt{N_{eff}}}
\end{equation}
and 
\begin{equation}
	\left \langle P_{coin} \right \rangle = \left \langle P \right \rangle^2 \left ( 1 + \frac{1}{N_{eff}} \right ).
\end{equation}

Assuming that each of the 36 beam crossings has the same luminosity, $L = L_{tot} / 36$, and that the background hits effectively arise from the previous $m$ beam crossings, we obtain
\begin{equation}
	N_\mathrm{eff} = m \frac{\sigma_\mathrm{LM} L}{f}
\end{equation}
\begin{figure}[h]
	\centering
	\subfigure{\label{fig:appB6a} \includegraphics[scale=0.47]{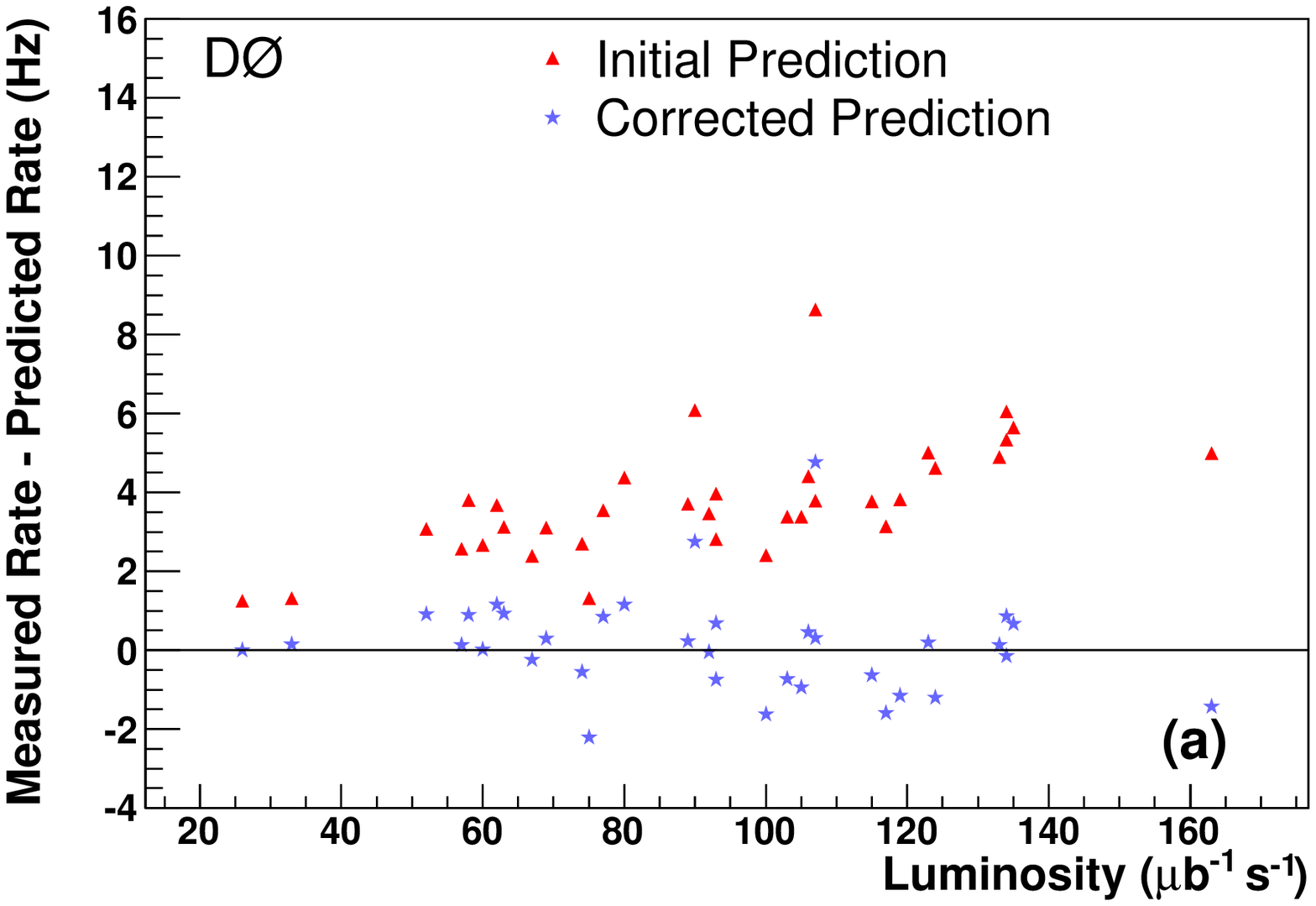}
	}
	\subfigure{\label{fig:appB6b} \includegraphics[scale=0.47]{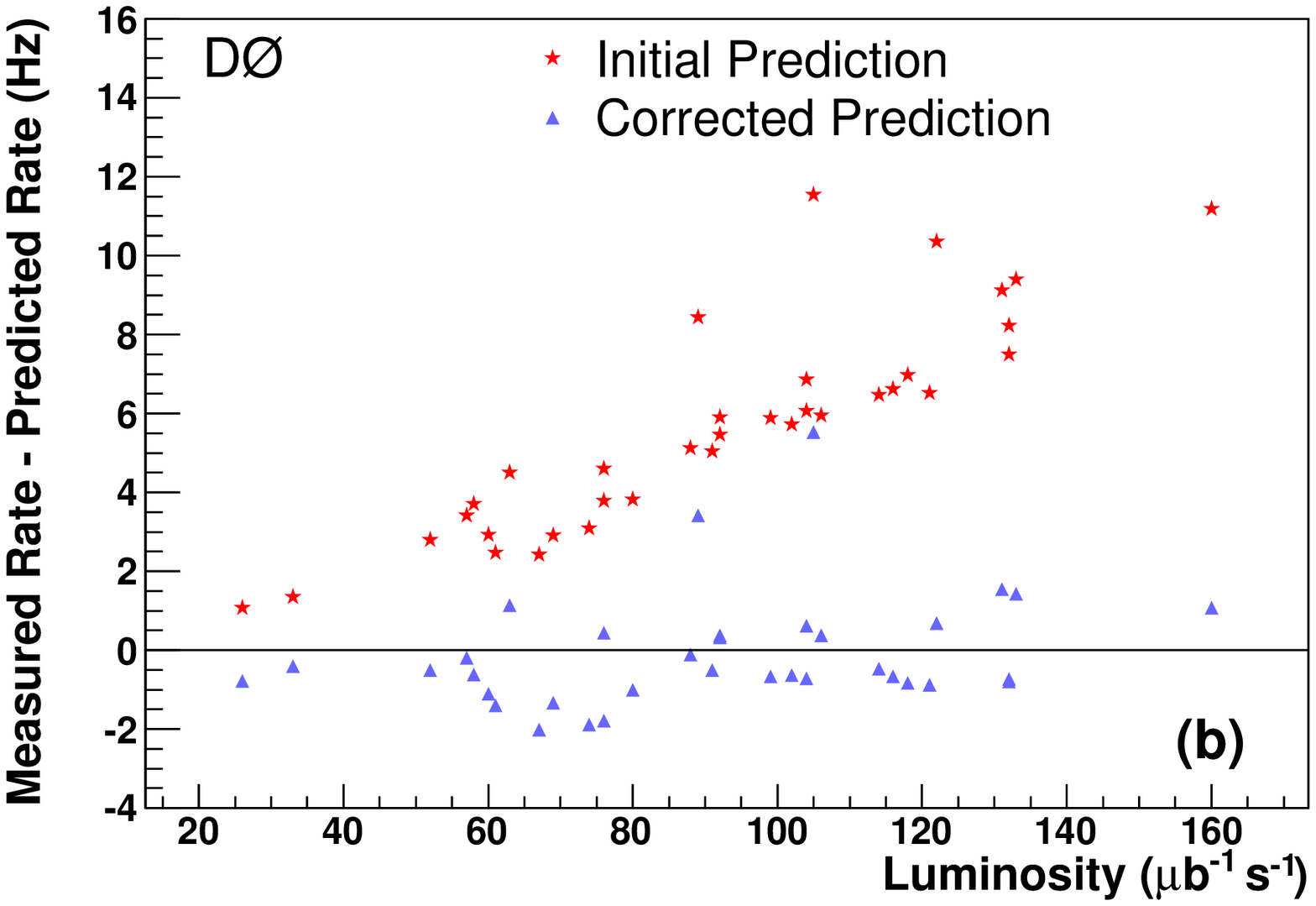}
	}
	\caption {Difference between the measured and predicted rates for bunch train crossings with hits in both north and south luminosity monitors (a) before and (b) after the bunch train. The prediction of Eq.~\ref{eq:n_ij} assumes a fixed background probability, whereas the corrected prediction allows the background probability to fluctuate.}
	\label{fig:appB6}
\end{figure}

The parameter $m$ is estimated by fitting the data. Using empty ticks prior to a bunch train gives an estimate of $m \approx 7$ and after the bunch train $m \approx 9$. The difference between the predicted and measured north -- south coincidence rates, both with and without correcting for fluctuations in the background probability, are shown in Fig.~\ref{fig:appB6}. 
The background model gives good agreement between the predicted and measured coincidence rate provided that the background probability is allowed to fluctuate.

\subsection{Background during Beam Crossings}

The aim is to estimate the background cross section for beam crossings, but the method discussed up to now is not applicable to actual beam crossings. A model is developed that allows the estimation of the background cross section during beam crossings. Using this model, it is found that the average background cross section for the 36 beam crossings is slightly higher than what is obtained by averaging the background cross sections before and after the bunch train.

The probability of producing a background hit in the north or south luminosity monitors is estimated from the coincidence probability
\begin{equation}
	P = \sqrt{\frac{P_{coin}}{1+ N^{-1}_{eff}} } ,
\end{equation}
\begin{figure}[h]
	\centering
	\includegraphics[scale=0.47]{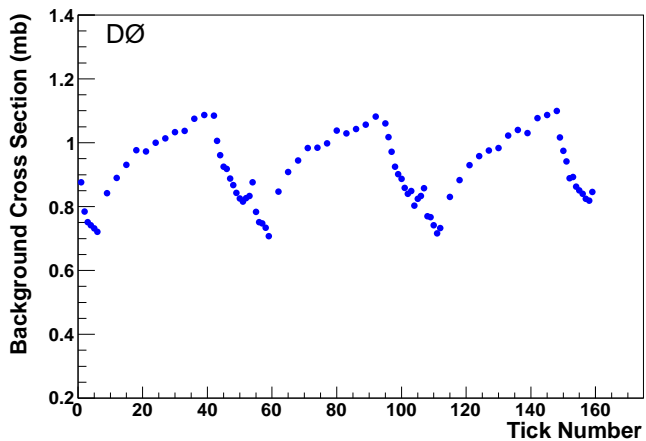}
	\caption {Background cross section during bunch crossings.}
	\label{fig:appB7}
\end{figure}

The corresponding background cross section is calculated using Poisson statistics to account for pileup
\begin{equation}
	\sigma_{BG} = - \frac{f}{L} \ln \left ( 1 - P \right ) . 
\end{equation}

Figure~\ref{fig:appB7} shows the background cross section for data recorded over the span of one minute on November 12, 2008 
at a luminosity of 68~$\mu$b$^{-1}$s$^{-1}$. The background cross section rises during the course of bunch crossings and then 
decays in the region without bunch crossings.
\begin{figure}[h]
	\centering
	\includegraphics[scale=0.47]{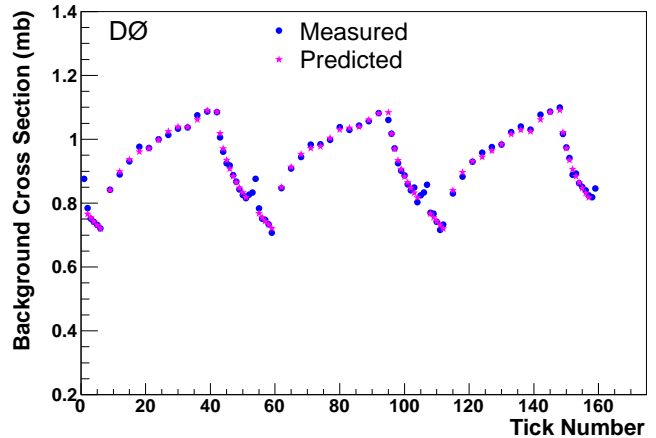}
	\caption {Comparison of the predicted background cross section and the measured background cross sections.}
	\label{fig:appB8}
\end{figure}

A model is constructed to describe the observed background cross section. The amplitude of the background produced by a given beam bunch is taken to be proportional to the luminosity for that bunch. It is assumed that the contribution to the background cross section falls exponentially in time following the bunch. There appears to be both a short and long time components in the background, thus a double exponential is used to fit the background. For a bunch occurring at $t = 0$ with luminosity $L$, the background model predicts that the contribution of the background to future ticks will be
%
\begin{equation}
	\sigma_{BG} (t) = L \left ( A_1 e^{-t / \tau_1} + A_2 e^{-t / \tau_2} \right ),
\end{equation}
%
where $A_1$ and $A_2$ are the amplitudes of the two background components and $\tau_1$ and $\tau_2$ are the associated time constants.
The background cross sections are fitted using this model. As shown in Fig.~\ref{fig:appB8}, 
this four parameter, two-component model provides a good description of the data. In this particular example, the gap region shows an anomalous up-tick in the background cross section that is likely due to residual beam in these ticks. In fitting the background model, the three ticks with more prominent anomalies have been excluded from the fit as well as the two preceding ticks.
\begin{figure}
	\centering
	\includegraphics[scale=0.47]{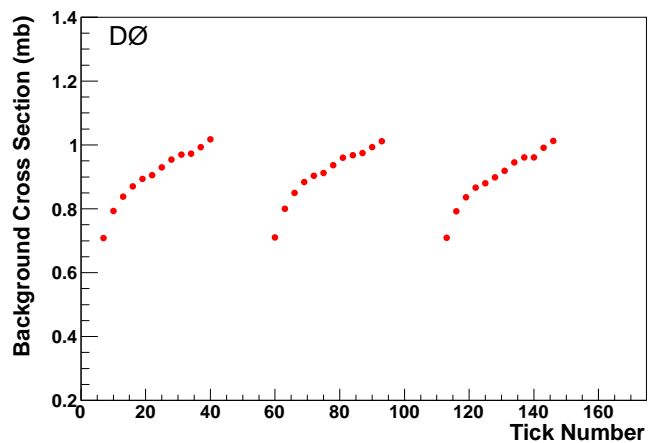}
	\caption {Predicted background cross section.}
	\label{fig:appB9}
\end{figure}

Since this model describes adequately the tick dependence of the background cross section, it is also used to estimate the background cross section during the beam crossings (see Fig.~\ref{fig:appB9}). The background cross section varies over time, thus the final background cross section estimate is based on the average over the 35 data samples used for this study. Previously (see Sec.~\ref{subsection:effective_background}), a background cross section of $ \sigma_{BG} =0.85$~mb was estimated by averaging the individual measurements for north and south monitors, before and after the bunch train. Correcting this value according to the model described above, a background cross section of $ \sigma_{BG} = 0.9 \pm 0.1$~mb is calculated, where the uncertainty quoted is the RMS spread.

\section{Background Subtraction}
	\label{app:bkgd_unfolding}
		The analytical method used to disentangle the background contribution from the measured multiplicity distributions is described below.
\subsection{Signal and Background Convolution}

Given $i$ counters with hits in one of the LM arrays, limits can be placed on the number of signal and background hits even though the same configuration of hits can be obtained from different combinations of signal and background hits. If $l$ counters have signal hits and $p$ counters have background hits, $l$ and $p$ will be constrained by $l \leq i$, $p \leq i$, and $l + p \geq i$. The last constraint is not an equality because there can be counters with both signal and background hits in them as multiple particles hitting a given counter are counted as a single hit.

The probability of having $i$ counters with hits can be constructed if the multiplicity distributions for signal and background are known. If $S_l$ and $B_p$ are the signal and background probabilities, then the probability $d_{ilp}$ for having $i$ observed counters with hits is given by
\begin{equation}
	d_{ilp} = S_l B_p f_{lpi},
\end{equation}
where $f_{lpi}$ is a combinatoric factor that gives the probability that $l$ counters with signal hits and $p$ counters with background hits will yield $i$ counters with observed hits. The combinatoric factor $f_{lpi}$ is derived in Section~\ref{sub:f_lpi}.

The total probability $D_i$ for observing $i$ hits is obtained by summing over the possible values of $l$ and $p$:
\begin{equation}
	D_i = \sum_{\substack{l = 0}}^{\substack{l \leq i }}
		  \sum_{\substack{p = 0}}^{\substack{p \leq i }}
		  S_i B_p f_{lpi} \Theta \left ( l + p - i \right ) , 
\end{equation}
where the constraints are explicitly represented through the summation limits and the use of the Heaviside step function $\Theta$:
\begin{equation}
		\Theta \left ( l + p - i \right ) = 
		\begin{cases}
			 1 & \mathrm{~for~} l + p \geq i , \\
			 0 & \mathrm{~for~} l + p < i .
		\end{cases} 
\end{equation}

Similarly, the probability $D_{ij}$ of having $i$ north counters and $j$ south counters with observed hits is given by
\begin{equation}
	\label{eq:probD}
	D_{ij}  = \sum_{\substack{l = 0 \\ m = 0}}^{\substack{l \leq i \\ m \leq j}}
		      \sum_{\substack{p = 0 \\ q = 0}}^{\substack{p \leq i \\ q \leq j}}
		      S_{lm} B_{pq} f_{lpi} f_{mqj} \Theta \left (l + p -i \right ) \Theta \left (m + q -j \right ) ,
\end{equation}
where $S_{lm}$ is the probability for having $l$ north counters and $m$ south counters with signal hits and $B_{pq}$ is the probability for having $p$ north counters and $q$ south counters with background hits.

\subsection{Derivation of Combinatoric Factor $f_{lpi}$}
	\label{sub:f_lpi}

The combinatoric factor $f_{lpi}$ gives the probability for observing $i$ counters with hits given $l$ counters with signal hits and $p$ counters with background hits. The procedure described is based on two assumptions:

\begin{enumerate}
	\item Background hits are uncorrelated with signal hits. The background hits originate from interactions in earlier beam crossings, while the signal hits originate from interactions in the current beam crossings, making this a reasonable assumption.
	\item The counters are either hit or not hit, ignoring the possibility that small amounts of charge from signal and background that are separately below threshold combine to exceed the discriminator threshold to form a hit. The same assumption is also made in the luminosity calculation and has been tested by comparing the north -- south coincidence rate in a large sample of $10^8$ simulated beam crossings with and without this assumption. No statistically significant difference was found.
\end{enumerate} 

If $u$ is the number of counters with both signal and background hits and $\nu$ is the number of counters with only background hits, we obtain
\begin{equation}
	\begin{split}
		& u = l + p - i , \\ 
		& \nu = i - l .
	\end{split}
\end{equation}

The number of ways to arrange $u$ hits with both signal and background among $l$ counters with signal hits is given by
\begin{equation}
	\label{eq:binom1}
	\binom{l}{u} = \frac{l!}{u! \left ( l - u \right )!} = \frac{l!}{ \left ( l + p - i \right )! \left ( i - p \right )! } .
\end{equation}
Similarly, the number of ways that $\nu$ background-only hits can be distributed among the $N-l$ counters without signal hits is
\begin{equation}
	\label{eq:binom2}
	\binom{N-l}{\nu} = \frac{ \left ( N - l \right )!}{\nu! \left (N - l - \nu \right )!} = \frac{ \left ( N - l \right )!}{ \left ( i - l \right )! \left ( N - i \right )! } .
\end{equation}
Thus, the total number of background hit combinations is the product of Eqs.~\ref{eq:binom1} and~\ref{eq:binom2}
\begin{equation}
	\label{eq:binomprod}
	\binom{l}{u}\binom{N-l}{\nu} = \frac{ l! \left ( N - l \right )!}{ \left ( l + p - i \right )! \left ( i - p \right )!  \left ( i - l \right )! \left ( N - i \right )! } ~.
\end{equation}
The total number of ways to arrange $p$ background hits among $N$ counters is
\begin{equation}
	\binom{N}{p} = \frac{N!}{p! \left ( N - p \right )!} ~.
\end{equation}
Each arrangement of background hits among the $N$ counters is assumed to be equally likely, therefore the probability of having $i$ counters with hits, given $l$ counters with signal hits and $p$ counters with background hits, is the number of arrangements meeting this condition divided by the total number of arrangements of the background hits:
\begin{equation}
	\begin{split}
		f_{lpi} & = \frac{ \binom{l}{u} \binom{N-l}{\nu} }{ \binom{N}{p} } \\
			  & = \frac{ l! \left ( N -l \right )! p! \left ( N - p \right )! }{ \left ( l + p -i \right )! \left ( i - p \right )! \left ( i - l \right )! \left ( N - i \right )!  N!} ~.
	\end{split}
\end{equation}
While the derivation did not treat signal and background hits symmetrically, the final result includes the expected $ l \leftrightarrow p$ symmetry.

\subsection{Unfolding Procedure}
	\label{sub:unfolding}

Once the data and background multiplicity distributions are measured, an unfolding procedure is used to extract the signal multiplicity distribution. 
To simplify the notation, the 2D multiplicity distributions, with indexes ranging from 0 to 24, are re-labeled to use a single index that ranges from 0 to 624. 
For example, the 2D multiplicity distributions for counters with observed hits can be written as:
\begin{equation}
	\begin{split}
		& D_\alpha \equiv D_{ij} \textrm{~~where~~} \alpha = 25i+j \\
		& S_\beta \equiv S_{lm} \textrm{~~where~~} \beta = 25l + m \\
		& B_\gamma \equiv B_{pq} \textrm{~~where~~} \gamma =25p+q.
	\end{split}
\end{equation}

The tensor $T_{\alpha \beta \gamma}$ is defined as follows:
\begin{equation}
	T_{\alpha \beta \gamma}  = 
	\begin{cases}
		f_{lpi} f_{jmq} & \textrm{~for~} \begin{split} & l \leq i, p\leq i, m \leq j, q \leq j, \\
		                				       			   & i \leq l + p, j \leq m + q \end{split} \\
		0                       & \textrm{~otherwise} .
	\end{cases}	
\end{equation}

In this notation, Eq.~\ref{eq:probD} for the convolution of signal and background multiplicity distributions can be written as
\begin{equation}
	\label{eq:tensor_conv}
	D_\alpha = T_{\alpha \beta \gamma} S_\beta B_\gamma ,
\end{equation}
where the convention of implied summation for repeated indices has been used.
To solve for the signal multiplicity distribution in this notation, the folding and unfolding matrices $F$ and $U$ are defined:
\begin{align}
	F_{\alpha \beta} & = T_{\alpha \beta \gamma} B_\gamma \\
	U & = F^{-1} .
\end{align} 
Substituting these matrices into Eq.~\ref{eq:tensor_conv} the following matrix equations are extracted:
\begin{align}
	D & = FS \\
	S & = F^{-1}D = UD
\end{align}

Thus, the signal multiplicity distribution is obtained by taking the product of the unfolding and data matrices. The inversion of the folding matrix $F$ is only necessary for a simultaneous determination of the covariance matrix for the signal multiplicity distribution. Otherwise a linear equation solver can be used to find the signal multiplicity distribution $S$.

\subsection{Covariance Matrix for the Signal Multiplicity Distribution}
	\label{sub:covmatrix}
	
The unfolding procedure introduces correlations among the multiplicity bins, so that the uncertainty is represented by a $625 \times 625$ element covariance matrix.
Standard error propagation techniques are used to determine the covariance matrix for the unfolded signal multiplicity distribution,
\begin{equation}
	\left ( \delta D \right )_\alpha = T_{\alpha \beta \gamma} \left ( \delta S \right )_\beta B_\gamma + T_{\alpha \beta \gamma}  S_\beta \left ( \delta B \right )_\gamma .
\end{equation}

This result can be formulated as a matrix equation:
\begin{align}
	G_{\alpha \gamma} & = T_{\alpha \beta \gamma} S_\beta \\
	\delta D & = F \left ( \delta S \right ) + G \left ( \delta B \right ) \\
	\delta S & = F^{-1} \left [ \left ( \delta D \right ) - G \left ( \delta B \right ) \right ] \nonumber \\
		     & = U \left [ \left ( \delta D \right ) - G \left ( \delta B \right ) \right ] .
\end{align}
The covariance matrix for the multiplicity distribution of counters with signal hits is given by:
\begin{align}
	C^S & \equiv \left \langle \left ( \delta S \right ) \left ( \delta S \right )^T \right \rangle \nonumber \\
		& = \left \langle U \left [ \left ( \delta D \right ) - G \left ( \delta B \right ) \right ]  \left [ \left ( \delta D \right ) - G \left ( \delta B \right ) \right ]^T U^T  \right \rangle .
\end{align}

Since the observed signal + background multiplicity distribution is uncorrelated with the background multiplicity distribution:
\begin{align}
	C^S & = U \left \langle \left ( \delta D \right ) \left ( \delta D \right )^T \right \rangle U^T + 
			UG \left \langle \left ( \delta B \right ) \left ( \delta B \right )^T \right \rangle G^TU^T \nonumber \\
		& = UC^DU^T + UGC^BG^TU^T , 
\end{align}
where $C^D$ and $C^B$ are the covariance matrices for the data and background multiplicity distributions.
The data and background covariance matrices simply contain diagonal binomial error terms:
\begin{align} 
	C^D_{\alpha \beta} & = \delta_{\alpha \beta} D_\alpha \left ( 1 - D_\alpha \right ) / N_D \\
	C^B_{\alpha \beta} & = \delta_{\alpha \beta} B_\alpha \left ( 1 - B_\alpha \right ) / N_B
\end{align}
where $\delta_{\alpha \beta}$ is the Kronecker $\delta$ function and $N_D$ and $N_B$ are the total number of entries in the data and background multiplicity distributions, respectively.

\end{document}